\documentclass[aps,prb,twocolumn,superscriptaddress]{revtex4-2}
\usepackage{graphicx}% Include figure files
\usepackage{amsmath}
\usepackage{xcolor}

\begin{document}

%\preprint{}

\title{Coupled dimerized alternating-bond quantum spin chains in the distorted honeycomb-lattice magnet Cu$_5$SbO$_6$}


%\thanks{}

\author{C.~Piyakulworawat}
\affiliation{Department of Physics, Faculty of Science, Mahidol University, Bangkok 10400, Thailand}
\affiliation{Thailand Center of Excellence in Physics, Commission of Higher Education, Bangkok 10400, Thailand}

\author{K.~Morita}
\affiliation{Department of Physics, Tohoku University, Sendai, Miyagi 980-8578, Japan}

\author{Y.~Fukumoto}
\affiliation{Department of Physics and Astronomy, Faculty of Science and Technology, Tokyo University of Science, Noda, Chiba 278-8510, Japan}

\author{W.~-Y.~Hsieh}
\affiliation{Center for Condensed Matter Sciences, National Taiwan University, Taipei 10617, Taiwan}
\affiliation{Taiwan Consortium of Emergent Crystalline Materials, National Science and Technology Council, Taipei 10622, Taiwan}

\author{W.~-T.~Chen}
\affiliation{Center for Condensed Matter Sciences, National Taiwan University, Taipei 10617, Taiwan}
\affiliation{Center of Atomic Initiative for New Materials, National Taiwan University, Taipei 10617, Taiwan}
\affiliation{Taiwan Consortium of Emergent Crystalline Materials, National Science and Technology Council, Taipei 10622, Taiwan}

\author{K.~Nakajima}
\affiliation{Materials and Life Science Division, J-PARC Center, Tokai, Ibaraki 319-1195, Japan}

\author{S.~Ohira-Kawamura}
\affiliation{Materials and Life Science Division, J-PARC Center, Tokai, Ibaraki 319-1195, Japan}

\author{Y.~Zhao}
\affiliation{Department of Materials Science and Engineering, University of Maryland, College Park, Maryland 20742, USA}
\affiliation{NIST Center for Neutron Research, National Institute of Standards and Technology, Gaithersburg, Maryland 20899, USA}

\author{S.~Wannapaiboon}
\affiliation{Synchrotron Light Research Institute, Muang, Nakhon Ratchasima 30000, Thailand}

\author{P.~Piyawongwatthana}
\affiliation{Materials and Life Science Division, J-PARC Center, Tokai, Ibaraki 319-1195, Japan}

\author{T.~J.~Sato}
\affiliation{Institute of Multidisciplinary Research for Advanced Materials, Tohoku University, Sendai, Miyagi 980-8577, Japan}
\affiliation{Neutron Science Laboratory, Institute for Solid State Physics, The University of Tokyo, Kashiwa, Chiba 277-8581, Japan}

\author{K.~Matan}
\email[]{Corresponding author: kittiwit.mat@mahidol.ac.th}
\affiliation{Department of Physics, Faculty of Science, Mahidol University, Bangkok 10400, Thailand}
\affiliation{Thailand Center of Excellence in Physics, Commission of Higher Education, Bangkok 10400, Thailand}

\date{\today}

\begin{abstract}

We analyze powder-averaged inelastic neutron scattering and magnetization data for the distorted honeycomb compound Cu$_5$SbO$_6$ using a first-order dimer expansion calculation and quantum Monte Carlo simulations. We show that, in contrast to the previously proposed honeycomb lattice model, Cu$_5$SbO$_6$ accommodates interacting dimerized spin chains with alternating ferromagnetic-antiferromagnetic couplings along the chain. Moreover, unlike the typical couplings observed in other Cu$^{2+}$-based distorted honeycomb magnets, the spin chains in Cu$_5$SbO$_6$ primarily couple through an antiferromagnetic  coupling that arises between the honeycomb layers, rather than the expected interchain coupling in the layers. This finding reveals a different magnetic coupling scheme for Cu$_5$SbO$_6$. In addition, utilizing x-ray spectroscopy and transmission electron microscopy, we also refine the crystal structure and stacking-fault model of the compound.

\end{abstract}

%\keywords{}

\maketitle

%\subsection{}
%\subsubsection{}

\section{Introduction}

%%Physics of dimerized magnets

Dimerized quantum magnets have constituted a fertile ground for exploring a plethora of exotic emergent quantum phenomena in magnetic insulators \cite{Sachdev2008, Vasiliev2018}. The most renowned aspect of such magnets is arguably quantum criticality. One of the quantum phase transitions is achieved in the field-induced Bose-Einstein condensation of magnons realized in several compounds \cite{Nikuni1999, Ruegg2003, Jaime2004, Zapf2006, Giamarchi2008, Aczel2009, Zapf2014, Samulon2009}. Phenomena in the vicinity of a quantum critical point separating a dimerized disordered phase from a long-range ordered phase, such as dimensional reduction \cite{Sebastian2006} and longitudinal excitation modes \cite{Ruegg2008}, have been observed. When frustration between exchange couplings among orthogonally oriented dimers are present as in the Shastry-Sutherland magnet, several other intriguing effects have been reported including topologically protected chiral edge modes of triplet excitations \cite{McClarty2017}, pressure-induced 4-spin plaquette singlet state \cite{Koga2000, Zayed2017}, field-induced spin-nematic phase \cite{Fogh2024}, incomplete devil's staircase \cite{Takigawa2013}, and crystallization of triplet bound states in magnetization plateaus \cite{Corboz2014}. Dimerized magnets are built upon spin dimers consisting of a pair of spins interacting with an intradimer antiferromagnetic (AFM) exchange coupling. The dimer, for the case of $S = \frac{1}{2}$, exhibits a nonmagnetic singlet ground state and degenerate triplet excited states in zero magnetic field, which are separated by a finite energy gap. Dimers that interact with each other with a ferromagnetic (FM) exchange coupling in a one-dimensional (1D) network, forming an alternating FM-AFM spin chain, form another distinct class of dimerized magnets. Magnetic materials that are characterized as quasi-1D FM-AFM dimerized spin chains include, as examples, copper nitrate \cite{Tennant2012}, BaCu$_2$V$_2$O$_8$ \cite{Klyushina2018}, CuNb$_2$O$_6$ \cite{Kodama1999}, (CH$_3$)$_2$NH$_2$CuCl$_3$ \cite{Stone2007}, etc.

%%NCSO, NCTO, LCSO, LCTO 

\begin{figure*}
	\includegraphics[width=\textwidth]{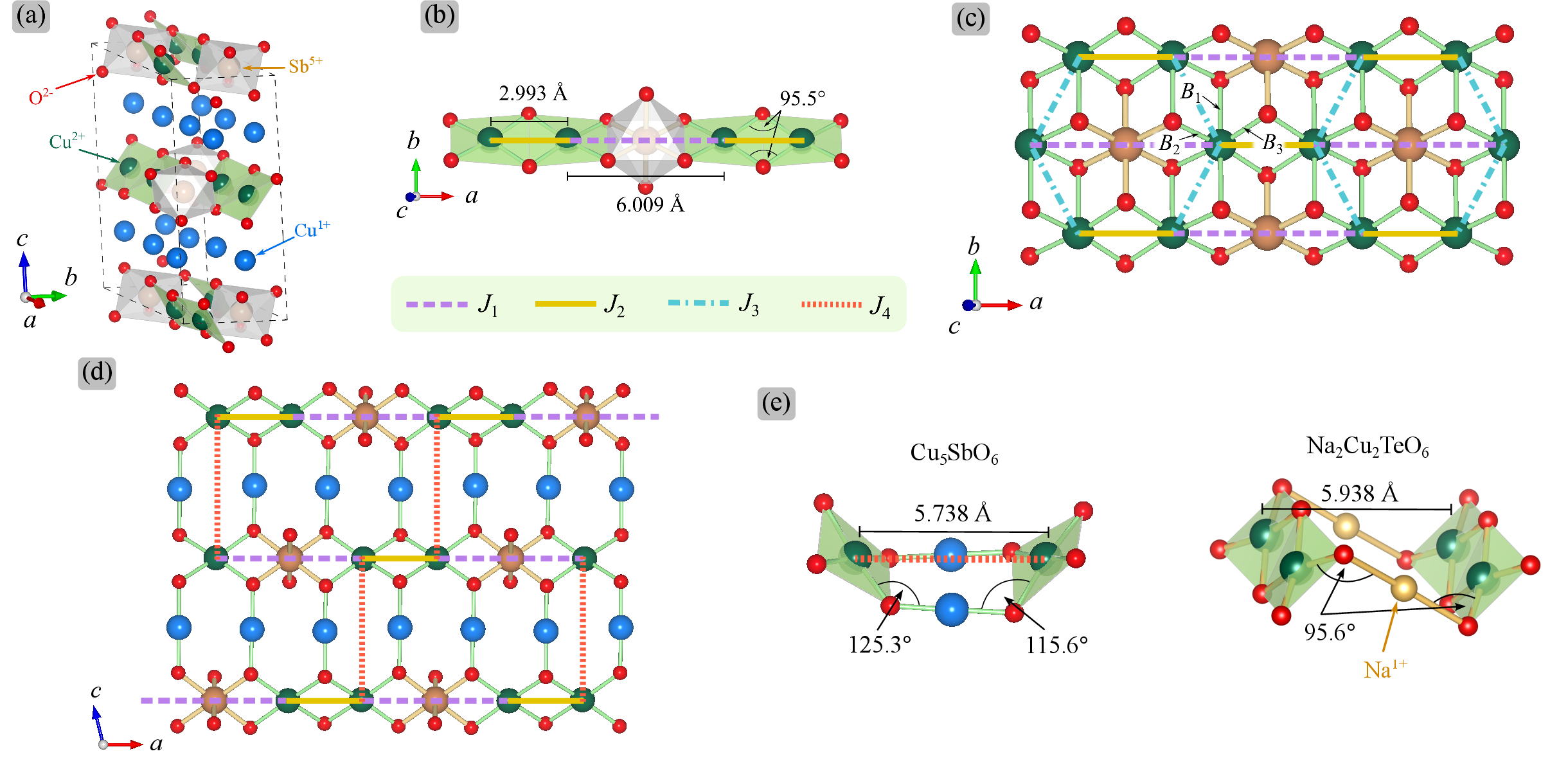}
	\caption{(a) A monoclinic unit cell of Cu$_5$SbO$_6$ is displayed in which spin chains consisted of SbO$_6$ octahedra (in silver) and edge-sharing Cu$_2$O$_6$ double plaquettes (in green) are aligned along the $a$ axis. The unit cell boundary is indicated by the dashed lines. (b) A chain segment with alternating $J_1$ and $J_2$ couplings is shown. The dashed purple and solid yellow lines represent intradimer $J_1$ and interdimer $J_2$ couplings, respectively. (c) The $ab$-plane projection of a single Cu$_2$SbO$_6$ layer is displayed. $B_{1}$ = 2.314 \AA\;is the Cu-O bond perpendicular to the plaquette whereas $B_{2}$ = 1.994 \AA\;and $B_{3}$ = 2.043 \AA\;are the Cu-O bonds within the plaquette. The dot-dash turquoise lines represent interchain $J_3$ couplings. (d) A single layer projected on to the $ac$ plane is shown, where the dotted orange lines represent interchain $J_4$ couplings perpendicular to the honeycomb planes. The super-exchange path responsible for the $J_4$ coupling in Cu$_5$SbO$_6$ is displayed in panel (e) along with the corresponding path in Na$_2$Cu$_2$TeO$_6$. The crystal structure illustrations are generated by \textsc{VESTA} \cite{Momma2011}.}
	\label{ucell}
\end{figure*}

To our knowledge, the archetypes of the quasi-1D $S = \frac{1}{2}$ alternating FM-AFM dimerized spin chain realized in \textit{a distorted honeycomb lattice} are Na$_3$Cu$_2$SbO$_6$ \cite{Miura2006, Miura2008, Koo2008, Kuo2012, Schmitt2014}, Li$_3$Cu$_2$SbO$_6$ \cite{Koo2016, Vavilova2021}, and Na$_2$Cu$_2$TeO$_6$ \cite{Koo2008, Schmitt2014, Gao2020, Shangguan2021, Lin2022}. It should be noted that, despite the similarity in their crystal structures, spin chains in Li$_3$Cu$_2$SbO$_6$ are fragmented due to a considerable site interchange between magnetic Cu$^{2+}$ and nonmagnetic Li$^{1+}$ ions \cite{Skakle1997, Nalbandyan2013, Koo2016, Vavilova2021}. We further note that a related compound Li$_2$Cu$_2$TeO$_6$ has also been synthesized \cite{Kumar2013}, but its magnetic characterization still awaits studies. The crystal structure of these compounds can be viewed as a stacking of Na$^{1+}$ layers alternating with Cu$_2$$M$O$_6$ layers where $M = \mathrm{Sb}, \mathrm{Te}$ \cite{Smirnova2005, Xu2005, Sankar2014, Murtaza2021}. A Cu$_2$$M$O$_6$ layer is built upon edge-sharing elongated CuO$_6$ octahedra forming a honeycomb lattice of Cu$^{2+}$ ions with $M$ ions positioned at the center of hexagons. The honeycomb lattice is distorted due to a strong Jahn-Teller effect on oxygen octahedra of Cu$^{2+}$ ions. Magnetic characteristics of these archetypal compounds are extensively studied by means of experiments as well as first-principles calculations. Experimental investigations include, for instance, inelastic neutron scattering (INS) \cite{Miura2008, Gao2020, Shangguan2021}, $^{23}$Na nuclear magnetic resonance (NMR) \cite{Morimoto2006, Kuo2012}, and thermodynamics measurements \cite{Xu2005, Miura2006, Derakhshan2007, Schmitt2014, Patil2025}. The experimental studies reveal a finite energy gap with the nonmagnetic singlet ground state in these systems. In particular, INS experiments unambiguously show that, at base temperatures, dimers are formed between third-nearest-neighbor rather than first-nearest-neighbor spins. Surprisingly, fitting the energy dispersion of the triplet excitations yields minimal AFM $J_3$ coupling and FM $J_2$ coupling that are weaker than AFM $J_1$ even though the bond length of $J_1$ is longer than that of $J_2$. Here, as indicated in Figs. \ref{ucell}(b) and \ref{ucell}(c), $J_2$ and $J_3$ denote the couplings between the first-nearest-neighbor spins whereas $J_1$ denotes the coupling between the third-nearest-neighbor spins.

First-principles calculations employing density functional theory (DFT) \cite{Xu2005, Derakhshan2007, Koo2008, Schmitt2014, Lin2022} and density-matrix renormalization group (DMRG) \cite{Lin2022} provide a comprehensive microscopic description for such a peculiar behavior of these exchange couplings deduced from experiments. The DFT results suggest that the magnetism of these compounds is primarily governed by the half-filled Cu$^{2+}$ $3d_{x^2 - y^2}$ orbital, where the local $x$ and $y$ axis almost align along Cu$^{2+}$-O$^{2-}$ bonds in the equatorial plane of the octahedron. This explains naturally the subordinate magnitude of $J_3$ coupling since the Cu$^{2+}$-O$^{2-}$-Cu$^{2+}$ path for $J_3$ incorporates a Cu$^{2+}$-O$^{2-}$ bond (bond $B_1$ in Fig. \ref{ucell}(c)) that is elongated and perpendicular to the Cu$^{2+}$ $3d_{x^2 - y^2}$ orbital. That the magnitude of $J_1$ coupling is much greater than that of $J_2$ coupling owes to the direct overlap of Cu$^{2+}$ $3d_{x^2 - y^2}$ orbitals with O$^{2-}$ $2p$ orbitals along the Cu$^{2+}$-O$^{2-}$-...-O$^{2-}$-Cu$^{2+}$ path across the $M$O$_6$ octahedron. On the other hand, as displayed in Fig. \ref{ucell}(b), since the angle of the Cu$^{2+}$-O$^{2-}$-Cu$^{2+}$ path for $J_2$ is close to 90$^\circ$, a pair of O$^{2-}$ $2p$ orbitals is almost orthogonal resulting in a weaker $J_2$. In accordance with Goodenough-Kanamori-Anderson rules \cite{Anderson1950, Goodenough1955, Goodenough1958, Moskvin1975}, the sign of $J_2$ coupling is expected to be FM.

%%CSO

Another compound in the family for exploring physics of the quasi-1D alternating dimerized spin chains has been reported in Ref. \cite{Climent2011}. The authors of Ref. \cite{Climent2011} have synthesized a Delafossite-derived compound Cu$_5$SbO$_6$ (Cu$^{1+}_3$Cu$^{2+}_2$SbO$_6$) that realizes a distorted honeycomb lattice of Cu$^{2+}$ spins. The crystal structure of Cu$_5$SbO$_6$ is built from a monoclinic unit cell, as shown in Fig. \ref{ucell}(a), with the space group $C2/c$ (no. 15) and the lattice parameters $a$ = 8.923 \AA, $b$ = 5.593 \AA, $c$ = 11.845 \AA, and $\beta$ = 103.585$^\circ$. The structure can also be viewed as a stacking of magnetic Cu$_2$SbO$_6$ layers alternating with nonmagnetic Cu$^{1+}$ layers. According to the measurements performed in Ref. \cite{Climent2011}, powder sample of Cu$_5$SbO$_6$ develops stacking faults, which manifest as diffusive streaks along the $c^*$ axis in a selected-area electron diffraction pattern. The stacking faults can be effectively improved by elevating synthesis temperature. The temperature dependence of magnetic susceptibility and magnetic heat capacity indicates a formation of spin dimers as temperature is lowered below 120 K. An approximated energy gap of $\sim$190 K is obtained by fitting susceptibility data with an isolated dimer model. While it is plausible to assume a similar $J_1$-$J_2$-$J_3$ magnetic coupling scheme for Cu$_5$SbO$_6$, a DFT calculation \cite{SaeFu2025} suggests a slightly different model. First of all, DFT calculation confirms the alternation of AFM $J_1$ and FM $J_2$ as in the case for related compounds. However, it is found that the AFM inter-plane coupling $J_4$ (see Fig. \ref{ucell}(d)) is significant, and $J_3$ is even smaller than in the related compounds. By inspecting the crystal structures of Cu$_5$SbO$_6$ and Na$_2$Cu$_2$TeO$_6$ in between the honeycomb planes, as shown in Fig. \ref{ucell}(e), one can see that the two angles Cu$^{2+}$-O$^{2-}$-Cu$^{1+}$ of the superexchange path Cu$^{2+}$-O$^{2-}$-Cu$^{1+}$-O$^{2-}$-Cu$^{2+}$ in Cu$_5$SbO$_6$ amount to about 120$^\circ$ which facilitates the electron exchange more efficiently between the two Cu$^{2+}$ ions compared to Na$_2$Cu$_2$TeO$_6$ where the angles Cu$^{2+}$-O$^{2-}$-Na$^{1+}$ are close to 90$^{\circ}$.

In this article, using neutron spectroscopy and magnetometry we set out to conclude that the $J_1$-$J_2$-$J_4$ model is more appropriate for Cu$_5$SbO$_6$. The content of this article is organized as follows. We first present studies of the crystal structure and stacking faults of the compound employing x-ray diffraction, x-ray absorption fine structure, and transmission electron microscopy techniques. Based upon the exchange coupling parameters determined from neutron data, quantum Monte Carlo simulations are performed to compute model temperature-dependent susceptibility and field-dependent magnetization, which are, in turn, used to verify with experimental data. Finally, we present an analysis of the powder-averaged inelastic neutron scattering data utilizing a first-order dimer expansion calculation for both $J_1$-$J_2$-$J_3$ and $J_1$-$J_2$-$J_4$ models.

\section{Experimental details}

\subsection{Crystal growth and characterization}

Powder samples of Cu$_5$SbO$_6$ were prepared by a conventional solid-state reaction from a stoichiometric mixture of CuO and Sb$_{2}$O$_{5}$. A homogeneous mixture of the precursors was placed in an alumina crucible and heated in air at 950 $^\circ$C for 24 h followed by an intermediate grinding and reheating at 1000 $^\circ$C for another 24 h. Single-crystal samples were grown from the obtained powder samples using the vertical-gradient freezing technique. The powder sample with mass $\sim$10 g was loaded into a pointed-bottom zirconia crucible and sealed with cement to prevent sample leakage during the growth process. We note that alumina crucibles do not work well with Cu$_5$SbO$_6$ due to sample penetration during crystallization from a liquid phase. The sealed crucible placed inside a quartz tube was positioned in a furnace where the temperature was $\sim$1160 $^\circ$C (the melting point) and uniform across the sample inside the crucible. The furnace was set up so that the temperature gradient below the starting point of the crucible was about $-$10 $^\circ$C/cm. The heating sequence started with increasing the temperature with a rate of 3 $^\circ$C/min from room temperature to 1160$^\circ$C. After the temperature reached 1160$^\circ$C, it was held for 6 h to ensure a complete and homogeneous melting. The sample was then lowered with a rate of 1 cm/day to start the crystallization. The process continued for 4 days before the temperature was decreased down to room temperature with a rate of 3 $^\circ$C/min.

The crystal structure of Cu$_5$SbO$_6$ at room temperature was probed with single-crystal and powder x-ray diffraction (XRD) techniques. The former employed a Bruker X8 APEX II CCD Diffractometer with Mo$K_\alpha$ radiation, whereas the latter used an Empyrean Diffractometer with Cu$K_\alpha$ radiation. A sample for powder diffraction was prepared by thoroughly grinding small pieces of crystals and loosely encased between Mylar C films. The powder XRD was performed in the transmission mode with rotating sample and with $10^\circ \le 2\theta \le 80^\circ$. The oxidation states of Cu ions at different crystallographic sites in Cu$_5$SbO$_6$ were determined by performing an x-ray absorption fine structure (XAFS) experiment. The experiment was conducted with the multi-x-ray techniques beamline BL1.1W at the Synchrotron Light Research Institute (SLRI) in Thailand. An XAFS spectrum of Cu$_5$SbO$_6$ powder at room temperature was obtained at the Cu$K$ edge with $8.8\;\mathrm{keV}\le E \le 9.8\;\mathrm{keV}$. To determine the edge energy of Cu$^{1+}$ and Cu$^{2+}$ ions, absorption spectra were taken separately from Cu$_2$O and CuO standards. The amplitude reduction factor of Cu atom was obtained from analyzing a Cu foil spectrum. A characterization of stacking faults was carried out by means of a transmission electron microscope (TEM) with powder samples synthesized at 1000 and 1150 $^\circ$C. The experiment was performed using a JEOL Electron Microscope equipped with a thermal field-emission electron gun and an in-column energy filter operating at 100 kV.

\subsection{Magnetometry and neutron spectroscopy}

Magnetization of Cu$_5$SbO$_6$ in an applied magnetic field oriented parallel ($\boldsymbol{H}_{ab}$) and perpendicular ($\boldsymbol{H}_{c^*}$) to the honeycomb plane was measured using a Quantum Design MPMS 3 vibrating-sample magnetometer. The temperature dependence was measured from $T = 2$ to 400 K with $H = 100$ Oe, whereas the field dependence was measured from $H = 0$ to $7\times10^4$ Oe at $T = 2$ and 120 K. For both field directions, an aligned crystal sample (see inset of Fig. \ref{xray}(a)) with mass 9.48~mg was attached to a quartz rod whose magnetization was measured separately under the same conditions.

Magnetic excitations were investigated with INS techniques. INS experiments in zero magnetic field were performed using the cold-neutron time-of-flight (TOF) disk-chopper spectrometer BL14 (AMATERAS) \cite{Nakajima2011} at Japan Proton Accelerator Research Complex and the double-focusing triple-axis (TA) spectrometer BT-7 \cite{Lynn2012} at NIST Center for Neutron Research. For both spectrometers, due to the unavailability of large single crystals, crystal samples with mosaicity were employed rather than genuine powder samples to effectively avoid stacking-fault issues. Powder averaging for the TOF data was done later using the \textsc{ustusemi} software \cite{Inamura2013}. At AMATERAS, a sample of mass $\sim2$ g was loaded in an aluminum can and cooled down to 4 K using a closed cycle $^4$He cryostat. The excitation spectra were probed with multiple neutron incident energies, including $E_\mathrm{i} = 23.7$ meV. The phonon spectrum was measured at 200 K with the same $E_\mathrm{i}$. During the measurement, the sample was rotated around the vertical axis with a step of $1^\circ$ to cover $172^\circ$. At BT-7, the experiment was performed using a sample of mass $\sim1$ g with the fixed-$E_\mathrm{f}$ mode having a final energy $E_\mathrm{f} = 14.7$ meV utilizing pyrolytic graphite monochromator and analyzer. The collimation adopted was open$-80^{\prime}-80^{\prime}-120^{\prime}$ with a single detector providing an energy resolution of 1.5 meV (FWHM) at the elastic position. The temperature was lowered to 2.8 K using a closed cycle $^4$He cryostat. The $E$-scan measurements were taken with a wave vector $Q \sim 2$ \AA$^{-1}$ in a temperature range $2.8\;\mathrm{K} \le T \le 200\;\mathrm{K}$, and at $T =$ 2.8 K the data were taken in a range $1 \;\mathrm{\AA}^{-1} \le Q \le 3\;\mathrm{\AA}^{-1}$.

\section{Results and Discussion}

\subsection{Crystal structure and stacking faults}

\begin{figure*}
	\includegraphics[width=\textwidth]{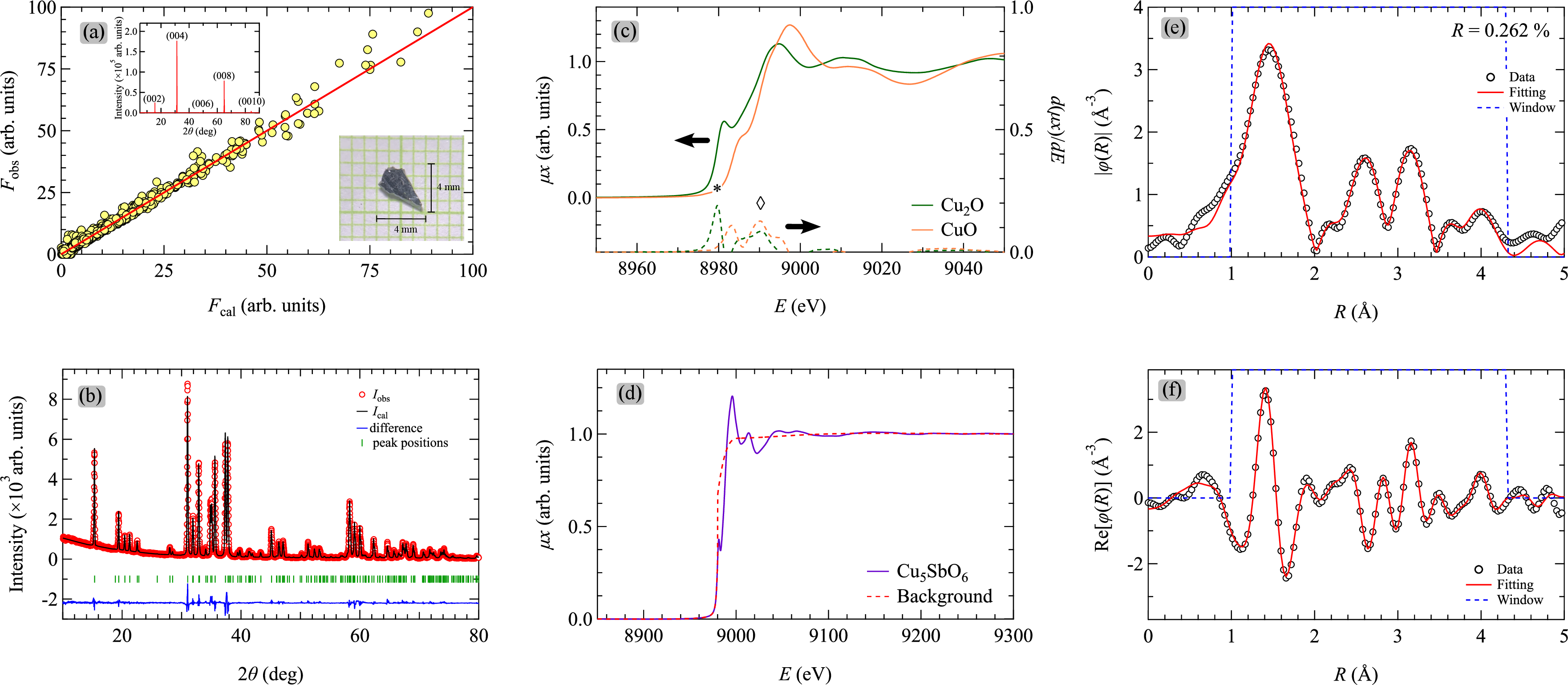}
	\caption{(a) A plot of $F_\mathrm{obs}$ against $F_\mathrm{cal}$ obtained from refinement of the single-crystal XRD data. The straight line is where $F_\mathrm{obs} = F_\mathrm{cal}$. The insets show a typical single-crystal sample and the $\theta$-$2\theta$ scan of a cleaved facet. (b) Powder XRD pattern of ground crystals. Open circles represent observed intensity. The black line represents the calculated intensity based on the refined parameters. The blue line shows the difference between observed and calculated intensities. The green bars mark the position of Bragg peaks. (c) Normalized absorption spectra of CuO (orange) and Cu$_2$O (green). The absorption coefficient $\mu x$ is shown on the left axis and the first derivative on the right axis. The asterisk and diamond symbols indicate the estimated edge energy of Cu$^{1+}$ and Cu$^{2+}$ ions, respectively. (d) Normalized absorption spectrum of Cu$_5$SbO$_6$. The dashed line represents background of the spectrum. (e) Modulus of the Fourier transform of $\varphi(k)$ with the real part shown in panel (f). The open circles represent the data, and the red curve is the fitting with the model described in text. The dotted line is the fitting range (window).}
	\label{xray}
\end{figure*}

\begin{figure*}
	\includegraphics[width=0.95\textwidth]{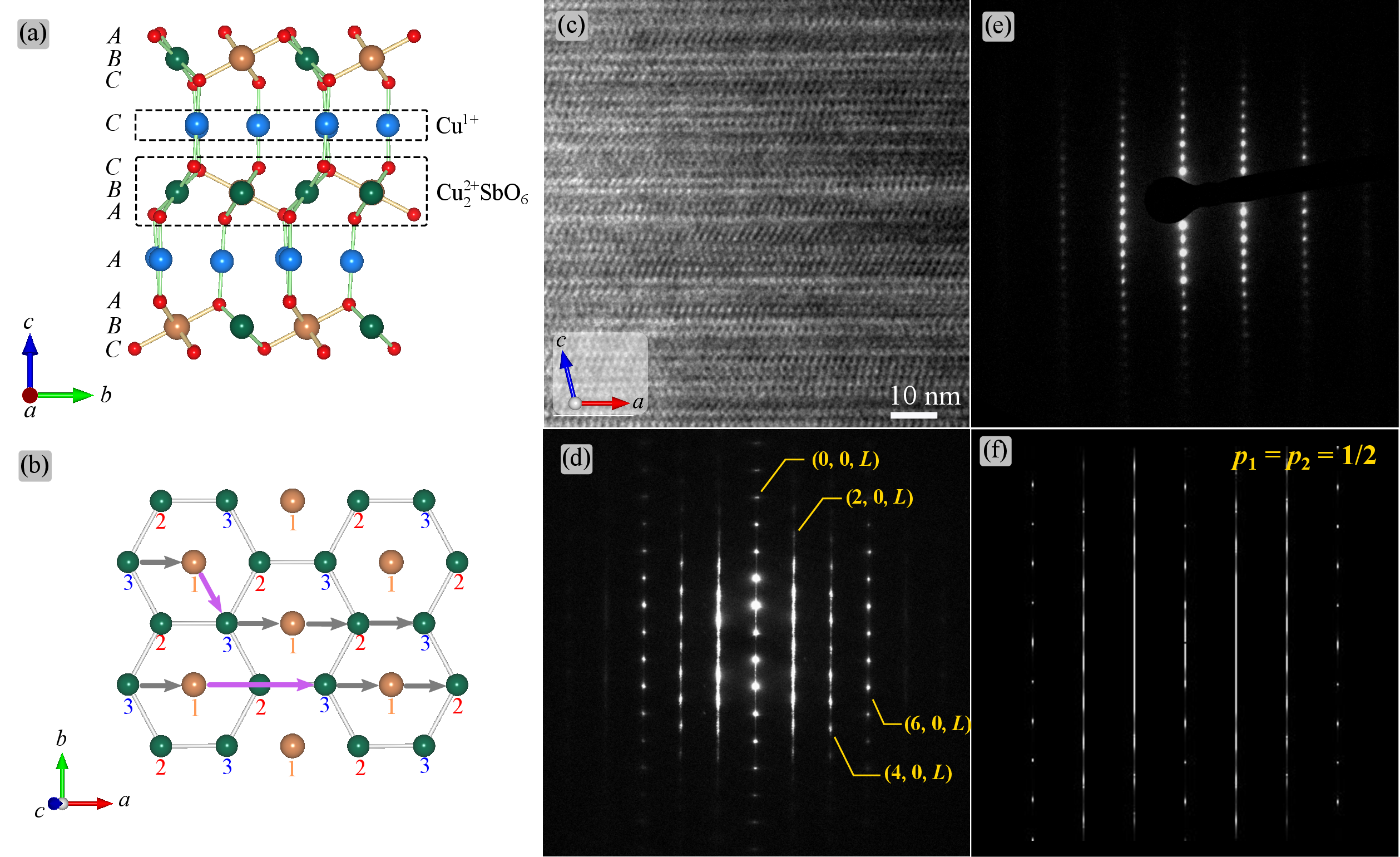}
	\caption{(a) A projection of the crystal structure onto the $bc^*$ plane showing the ideal stacking sequence of Cu$_5$SbO$_6$. (b) Two equivalent representatives of in-plane translations of the honeycomb layer that contain one stacking fault depicted by the purple arrows. The gray arrows represent the ideal translation vector. ``1," ``2," and ``3" are sublattice labels of the honeycomb layer. The atom representations are the same as in Fig. \ref{ucell}(a). (c) A TEM image in the $ac$ plane of the sample synthesized at 1000 $^\circ$C. (d), (e) SAED patterns in the $(H, 0, L)$ planes of samples synthesized at 1000 and 1150 $^\circ$C, respectively. (f) A simulated SAED pattern computed with $p_1 = p_2 = 1/2$. The crystal structure illustrations are generated by \textsc{VESTA} \cite{Momma2011}.}
	\label{tem}
\end{figure*}

\begin{table}
\caption{\label{refine}Optimized atomic parameters of Cu$_5$SbO$_6$ obtained from refinements of the single-crystal and powder XRD data taken at room temperature. The unit cell is monoclinic with the space group $C2/c$ (no. 15). The lattice parameters are $a$ = 8.9291(2) \AA, $b$ = 5.5974(1) \AA, $c$ = 11.8529(2) \AA, and $\beta$ = 103.605(1)$^\circ$. Numbers in parentheses are one standard deviation.}
	\begin{ruledtabular}
	\begin{tabular}{ccccc}
	Atom & Site & $x/a$ & $y/b$ & $z/c$ \\
	\hline
	\multicolumn{5}{c}{Single-crystal x-ray diffraction} \\
	Sb & $4c$ & 3/4 & 3/4 & 0 \\
	Cu(1) & $4e$ & 0 & 0.6243(3) & 1/4 \\
	Cu(2) & $8f$ & 0.8285(1) & 0.0968(3) & 0.2437(1) \\
	Cu(3) & $8f$ & 0.4168(1) & 0.7487(2) & 0.00175(9) \\
	O(1) & $8f$ & 0.4451(8) & 0.118(1) & 0.0912(6) \\
	O(2) & $8f$ & 0.2227(8) & 0.924(1) & 0.9178(6) \\
	O(3) & $8f$ & 0.6176(8) & 0.381(1) & 0.5969(6) \\
	\multicolumn{5}{c}{$R = 6.44\%$, w$R_2 = 17.7\%$, GoF = 2.22} \\
	\hline
	\multicolumn{5}{c}{Powder x-ray diffraction} \\
	Sb & $4c$ & 3/4 & 3/4 & 0 \\
	Cu(1) & $4e$ & 0 & 0.6251(6) & 1/4 \\
	Cu(2) & $8f$ & 0.8285(4) & 0.0969(4) & 0.2427(3) \\
	Cu(3) & $8f$ & 0.4164(3) & 0.749(1) & 0.0019(3) \\
	O(1) & $8f$ & 0.450(1) & 0.112(2) & 0.092(1) \\
	O(2) & $8f$ & 0.225(1) & 0.920(2) & 0.916(1) \\
	O(3) & $8f$ & 0.615(1) & 0.377(2) & 0.595(1) \\
	\multicolumn{5}{c}{$R_\mathrm{p} = 6.42\%$, $R_\mathrm{wp} = 9.05\%$, GoF = 1.87}
	\end{tabular}
	\end{ruledtabular}
\end{table}

Refinements of the crystal structure using the collected single-crystal and powder XRD data are based on the structural model reported in Ref. \cite{Climent2011}. The refinement with the single-crystal XRD data is performed with 799 unique reflections with $I_\mathrm{obs} > 3\sigma$, where $I_\mathrm{obs}$ and $\sigma$ are the observed intensity and corresponding standard deviation, respectively, using \textsc{jana2020} software package \cite{Petricek2014}. The refinement result is presented in Fig. \ref{xray}(a) as a plot of observed structure factor $F_\mathrm{obs}$ against calculated structure factor $F_\mathrm{cal}$ computed from the refined atomic parameters listed in Table \ref{refine}. As can be seen in Fig. \ref{xray}(a), the model provides the values of the calculated structure factor that are fairly consistent with the observed data. The cleavage of a typical crystal was examined with the $\theta$-$2\theta$ scan and was found to be $(00L)$ planes as shown in the inset of Fig. \ref{xray}(a). Figure \ref{xray}(b) displays the powder XRD pattern obtained from ground crystals. Comparing the position of the observed Bragg peaks with that derived from the model (green bars) indicates that impurities in the sample, if present, are negligible. The diffraction pattern is fitted to the model using the Rietveld method \cite{Rietveld1969} as implemented in \textsc{fullprof} software suite \cite{Rodriguez-Carvajal1993}. The refinement yields the optimized lattice and atomic parameters listed in Table \ref{refine}. The calculated diffraction pattern generated from the so-obtained parameters is illustrated in Fig. \ref{xray}(b) as a black solid line, which fits the observed data fairly well.

%%Cu$^{1+}$ and Cu$^{2+}$ sites

While XRD is one of the most powerful techniques for investigating the periodic structure of compounds, it is, however, practically incapable of distinguishing ions that have similar atomic factors as in the case of Cu$^{1+}$ and Cu$^{2+}$ in Cu$_5$SbO$_6$. Namely, while coordinates of Cu(1), Cu(2) (between honeycomb planes), and Cu(3) (in honeycomb planes) in the unit cell can be accurately identified, their oxidation states must be determined by another method. As Cu$^{1+}$ and Cu$^{2+}$ are quite similar in size, it is possible to have site interchange between them as in the case of Li$_3$Cu$_2$SbO$_6$ in which considerable amounts of Li$^{1+}$ and Cu$^{2+}$ ions exchange sites \cite{Koo2016, Vavilova2021}. To confirm that all Cu(3) in Cu$_5$SbO$_6$ has 2+ oxidation state and, therefore, form full Cu$^{2+}$ honeycomb layers, we turn to XAFS technique, which is more effective for determining the oxidation state of a specific ion. The absorption coefficient, $\mu x$, of a material can be obtained from $I = I_0e^{-\mu x} \Leftrightarrow \mu x = -\ln (I/I_0)$, where $I$ and $I_0$ are transmitted and incident intensities, respectively. Figures \ref{xray}(c) and \ref{xray}(d) show the collected XAFS spectra of CuO, Cu$_2$O, and Cu$_5$SbO$_6$, respectively. Inspection on the results in Fig. \ref{xray}(c) shows that Cu$^{1+}$ ions in Cu$_2$O (green) and Cu$^{2+}$ ions in CuO (orange) have unequal edge energies. Estimated values of edge $E_0$ for Cu$^{1+}$ and Cu$^{2+}$ ions are determined from their absorption spectra as the energy $E$ at which the first derivative of $\mu x$ is maximized, which is done using the \textsc{athena} module in the \textsc{demeter} software suite \cite{Ravel2005}. The first derivatives of $\mu x(E)$ for both Cu$_2$O and CuO are displayed as dashed lines in Fig. \ref{xray}(c) with the asterisk and diamond symbols marking the estimated $E_0$ of Cu$^{1+}$ and Cu$^{2+}$ ions, respectively. The difference in $E_0$ of the two ions is about 10 eV.

To determine the oxidation states of Cu ions in Cu$_5$SbO$_6$, we performed the extended x-ray absorption fine structure (EXAFS) analysis. EXAFS analysis is performed in the oscillating region of the spectrum \cite{Newville2014}, which is quantified by $\varphi(k) \equiv \mu x - \mu_0 x$, where $\mu_0 x$ is the background signal (dashed line in Fig. \ref{xray}(d)) and $k$ is related to $E$ through $k = \sqrt{2m_\mathrm{e}(E - E_0)}/\hbar$. Here, $m_\mathrm{e}$ is the electron mass. Contrary to XRD that depends upon average periodic structures, EXAFS relies on local environments around a particular ion. As the local environment of Cu sites within the honeycomb planes and that in between the planes are completely different, each site, therefore, produces a distinct oscillating pattern in $\varphi(k)$. To construct a model for the fitting, the crystal structure listed in Table \ref{refine} is used to generate all possible electron scattering paths around each Cu site, and $\varphi(k)$ of Cu$_5$SbO$_6$ is then a superposition of $\varphi_i(k)$s from paths having significant amplitudes. To facilitate the fitting, the observed spectrum is Fourier-transformed to a real-space variable $R$, whose modulus and real part are illustrated in Figs. \ref{xray}(e) and \ref{xray}(f), respectively. The fitting is performed using the \textsc{artemis} module in the \textsc{demeter} software suite \cite{Ravel2005}. The fitting parameters include path distance corrections $\delta R$ along with thermal deviations $\sigma$, Cu amplitude reduction factor $S_0^2$, and edge corrections $\delta E_0$. The amplitude reduction factor $S_0^2$ for Cu atom is obtained separately from fitting $\varphi(R)$ collected with a Cu foil sample and is found to be 0.89(5). The initial values of $E_0$ for each Cu site are set to the value of Cu$^{1+}$ ion, namely, the value marked by the asterisk in Fig. \ref{xray}(c). The fitting result is shown as red curves in Figs. \ref{xray}(e) and \ref{xray}(f) with $R = 0.262\%$. The optimized values of $\delta E_0$ for Cu(1) and Cu(2) are found to be the same (and, therefore, are constrained to the same value in the final refinement), which equals to 3.0(5) eV, while $\delta E_0$ of Cu(3) is 13.2(3) eV. Hence, $E_0$ of Cu(3) residing in the honeycomb planes is larger than $E_0$ of the other two sites in between the planes by about 10 eV, i.e., the oxidation state of Cu(3) is 2+ and that of Cu(1) and Cu(2) is 1+.

%%Stacking faults

As a common feature of several layered compounds, stacking faults, a consequence of multiple in-plane translations that virtually contribute the same amount of energy to the crystal, also occur in Cu$_5$SbO$_6$ \cite{Climent2011}. A precise description of such a crystallographic defect is crucial as it can potentially affect the magnetic coupling scheme of the compound. To appreciate how stacking faults form in Cu$_5$SbO$_6$, let us first investigate the ideal stacking as depicted in Figs. \ref{tem}(a) and \ref{tem}(b), and the stacking-fault model is developed by following the strategy discussed in Refs. \cite{Breger2005, Choi2012}. A projection of the crystal structure onto the $bc^*$ plane in Fig. \ref{tem}(a) shows that the ideal stacking is an alternation of Cu$^{2+}_2$SbO$_6$ slabs and Cu$^{1+}$ layers. A Cu$^{2+}_2$SbO$_6$ slab consists of a Cu$^{2+}$ honeycomb layer with Sb$^{5+}$ at the center of each hexagon enclosed by two O$^{2-}$ layers in the $ABC$ close-pack manner. The alternation results in a recursive $AAABCCCB$ sequence in which the Cu$^{2+}$ honeycomb layer is located at the $B$ layer. As depicted in Fig. \ref{tem}(b), a Cu$^{2+}$ honeycomb layer can be decomposed into three sublattices denoted by ``1," ``2," and ``3" in which Sb$^{5+}$ ions reside in sublattice ``1" and Cu$^{2+}$ ions in sublattices ``2" and ``3." Inspecting Fig. \ref{ucell}(d) shows that the ideal stacking corresponds to in-plane translations of Sb$^{5+}$ ions in a recursive ``123" sequence as shown in Fig. \ref{tem}(b) by gray arrows.

A selected-area electron diffraction (SAED) pattern for the powder sample synthesized at 1000 $^\circ$C is shown in Fig. \ref{tem}(d) in which diffusive scattering streaks are observed along $(2, 0, L)$ and $(4, 0 , L)$ but are absent along $(0, 0, L)$ and $(6, 0 , L)$. The streaks observed reflect lack of periodicity in the $c^*$ direction pointing to translations of the Cu$^{2+}$ honeycomb layers that intermittently intervene the ideal ``123" sequence. To gain more insight into this mistranslation, we next investigate a TEM image of the $ac$ plane of this sample as displayed in Fig. \ref{tem}(c). The image clearly features an irregularity in the stacking pattern and hints at the presence of two in-plane translations along the $a$ direction. Inspecting Fig. \ref{tem}(b) suggests that, in addition to the ideal in-plane translation, there is another possible translation (depicted by purple arrows) of the Cu$^{2+}$ honeycomb layers that translates Sb$^{5+}$ from sub-lattice ``1" to ``3," ``3" to ``2," and ``2" to ``1." Thus, an appropriate model for stacking faults in Cu$_5$SbO$_6$ corresponds to intervening the ideal ``123" sequence by a stacking-fault translation at certain frequency. It is worth noting that the ideal in-plane translation vector can be represented by (1/3, 0, 0) and that of the stacking fault by $(2/3, 0, 0)$ or, equivalently, $(-1/3, 0, 0)$. This model immediately provides a description for the disappearance of diffusive streaks along $(6, 0, L)$. As the stacking-fault translation vector is $(2/3, 0, 0)$ that can be obtained by adding $(1/3, 0, 0)$ to the ideal translation vector. The additional phase factor caused by the extra $(1/3, 0, 0)$ results in a phase difference of $2\pi$ for $(6, 0, L)$ planes, hence, constructive interference. The streaks are not observed along $(0, 0, L)$ since the phase factor due to these planes always adds in-phase irrespective of the in-plane translations. To further verify the model, a SAED pattern is simulated using \textsc{faults} software package \cite{Treacy1991, Casas-Cabanas2006, Casas-Cabanas2016}. The frequency of the ideal $p_1$ and of the stacking-fault translations $p_2$ are introduced, which are subjected to a constraint $p_1 + p_2 = 1$. Figure \ref{tem}(f) illustrates a simulated SAED pattern of an extreme case where $p_1 = p_2 = 1/2$ which can well reproduce the diffusive intensities along $(2, 0 , L)$ and $(4, 0, L)$ in the observed SAED pattern in Fig. \ref{tem}(d). In addition, the model can also reproduce the broad peak in the XRD data of a powder sample as illustrated in the inset of Fig. 5(a) of Ref. \cite{Climent2011}.

A SAED pattern for the sample synthesized at 1150~$^\circ$C, a temperature just below the melting point, shows a strikingly different result as shown in Fig. \ref{tem}(e). In this case, the diffusive scattering streaks are unobserved reflecting a restoration of the periodicity along the $c^*$ direction. The absence of the diffusive streaks in this SAED pattern is consistent with the absence of peak broadening, especially in $19^\circ < 2\theta < 23^\circ$, in the powder XRD data collected from ground crystals (see Fig. \ref{xray}(b)). The improvement of stacking faults in Cu$_5$SbO$_6$ by raising the synthesis temperature was also reported in Ref. \cite{Climent2011}.

\subsection{Magnetization}

\begin{figure}
	\includegraphics[width=0.9\columnwidth]{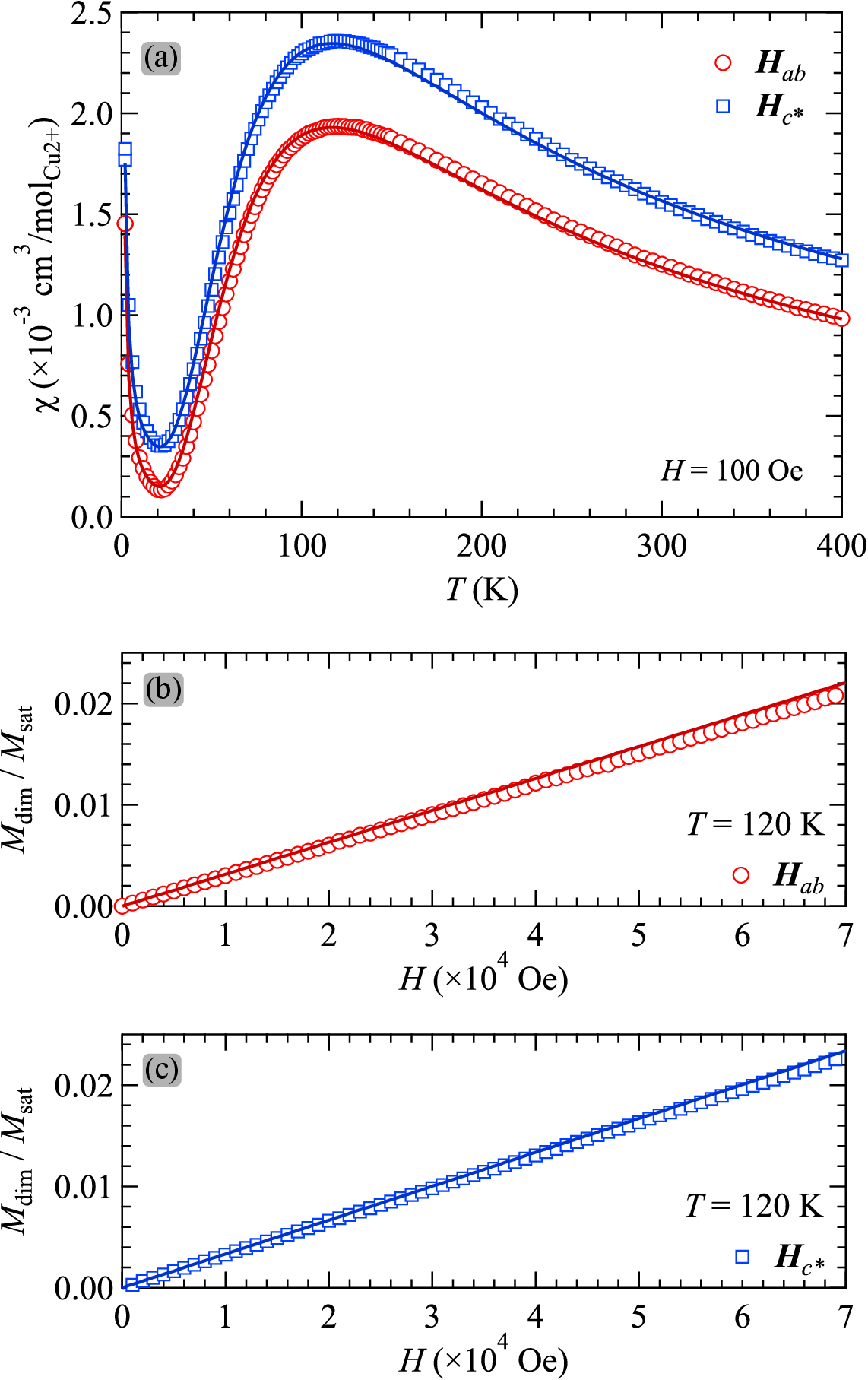}
	\caption{(a) Magnetic susceptibility as a function of temperature, $\chi(T)$, measured with an applied magnetic field $H = 100$ Oe. The solid curves are the best fits. (b), (c) Magnetization of dimerized spins $M_\mathrm{dim}$ as a function of applied magnetic field normalized with the saturated magnetization $M_\mathrm{sat} = g\mu_\mathrm{B}S$ measured at $T = 120$ K. The solid lines are model calculations described in text. The data are collected with $\boldsymbol{H}$ parallel to the $ab$ plane (circle symbols) and to the $c^*$ axis (square symbols).}
	\label{chi}
\end{figure}

The temperature dependence of magnetic susceptibility, $\chi = M/H$, for the in-plane ($\boldsymbol{H}_{ab}$) and out-of-plane ($\boldsymbol{H}_{c^*}$) field directions are shown in Fig. \ref{chi}(a). The holder's susceptibility measured separately is found to be smaller than the measurement uncertainty of the sample's susceptibility and is, therefore, neglected. As the temperature decreases from 400 K, a broad peak  indicative of short-range correlations develops at 120 K followed by an upturn below 20 K. No signature of thermal phase transitions is observed down to 2 K. Regardless of the upturn, the absence of anomalies along with the gradual decrease in $\chi(T)$ toward zero suggests the formation of spin dimers in Cu$_5$SbO$_6$ at low temperatures. To further investigate the upturn below 20 K, magnetization curves, $M(H)$, for both field directions were measured at 2 K where the signal from dimerized spins can be safely ignored. It is found that the $M(H)$ curves exhibit a Brillouin function that resembles Curie paramagnetism and, therefore, points to the presence of quasi-free spins in the sample under study. Hence, an appropriate model of the susceptibility can be expressed as
\begin{align}
	\chi(T) = f_\mathrm{free}\chi_\mathrm{free}(T) + f_\mathrm{dim}\chi_\mathrm{dim}(T) + \chi_0
	\label{model_chi}
\end{align}
where $f_\mathrm{free}$ and $f_\mathrm{dim}$ are population fractions of quasi-free and dimerized spins, respectively, with a constraint $f_\mathrm{free} + f_\mathrm{dim} = 1$. $\chi_0$ is a temperature-independent background susceptibility from other nonmagnetic ions in Cu$_5$SbO$_6$ and core shells of Cu$^{2+}$. $\chi_\mathrm{free}(T) = C/T$ with the Curie constant expressed as $C = N_\mathrm{A}p^2\mu^2_\mathrm{B}/(3k_\mathrm{B})$, where $N_\mathrm{A}$ is Avogadro constant, $\mu_\mathrm{B}$ is Bohr magneton, $k_\mathrm{B}$ is Boltzmann constant, and $p = g\sqrt{S(S + 1)}$ is the effective Bohr magneton number with $g$ being Landé factor of quasi-free spins.

We determine $C$ and $f_\mathrm{free}$ from the $M(H)$ curves for both field directions measured at $T = 2$ K. To model the $M(H)$ curve at this temperature, two terms responsible for paramagnetic and background magnetization are required. These two terms are $M_\mathrm{free}(H) \propto B_{1/2}(H)$, where $B_{1/2}(H)$ is a Brillouin function for $S = \frac{1}{2}$, and $M_0(H) \propto H$. Therefore, 
\begin{align}
	M(H) = \frac{1}{2}g\mu_\mathrm{B}N_\mathrm{free}\tanh\left(\frac{g\mu_\mathrm{B}H}{2k_\mathrm{B}T}\right) + M_0(H),
	\label{brill}
\end{align}
where $N_\mathrm{free} = f_\mathrm{free}N_\mathrm{tot}$ and $N_\mathrm{free}$, $N_\mathrm{tot}$ are the number of quasi-free and total spins, respectively, in the sample used. By fitting Eq.(\ref{brill}) to the $M(H)$ data at 2 K, we can determine $f_\mathrm{free} = 0.00684(6)$ and $C = 0.461(5)$ cm$^3$mol$^{-1}_\mathrm{Cu^{2+}}$K for $\boldsymbol{H}_{c^*}$ and $f_\mathrm{free} = 0.00608(4)$ and $C = 0.457(4)$ cm$^3$mol$^{-1}_\mathrm{Cu^{2+}}$K for $\boldsymbol{H}_{ab}$.

We perform quantum Monte Carlo (QMC) simulations to acquire a model for $\chi_\mathrm{dim}(T)$. The reduced susceptibility, $\chi^*(t)$, for $0.01 \le t \le 10$, where $t = k_\mathrm{B}T/J_\mathrm{max}$ is calculated using the \textsc{loop} algorithm \cite{Todo2001, Evertz2003} implemented in the \textsc{alps} simulation package \cite{Bauer2011}. The simulation is performed for the $J_1$-$J_2$-$J_4$ model with $J_\mathrm{max} = J_1 = 16.5$~meV, $J_2 = -6$~meV, and $J_4 = 3$~meV. The determination of these coupling parameters will be discussed in the next section. The simulations are performed on a cluster of 100$\times$100 spins with a periodic boundary condition using 100~000 Monte Carlo steps for thermalization and 500~000 Monte Carlo steps after thermalization. To obtain a functional form of $\chi^*(t)$, we fit simulation results with \cite{Johnston2000}
\begin{align}
   \chi^*(t) = \frac{\exp(-\Delta^*/t)}{4t}\mathcal{P}^{(q)}_{(r)}(t),
   \label{pade}
\end{align}
where $\mathcal{P}^{(q)}_{(r)}(t)$ is the Padé approximant expressed as
\begin{align*}
	\mathcal{P}^{(q)}_{(r)}(t) = \frac{1 + \sum_{n = 1}^{q}N_n/t^n}{1 + \sum_{n = 1}^{r}D_n/t^n}.
\end{align*}
Here $\Delta^*$, $N_n$, and $D_n$ are fitting parameters. We have found that the Padé approximant with $q = r = 5$, i.e., $\mathcal{P}^{(5)}_{(5)}(t)$, yields a sufficient accuracy with the deviation between the simulation results and Eq. (\ref{pade}) in the order of $10^{-3}$. $\chi_\mathrm{dim}(T)$ is related to $\chi^*(t)$ by
\begin{align}
   \chi_\mathrm{dim}(T) = \frac{N_\mathrm{A}g^2\mu_\mathrm{B}^2}{J_\mathrm{max}}\chi^*\left(\frac{k_\mathrm{B}T}{J_\mathrm{max}}\right).
   \label{chi_dim}
\end{align}
We fit the experimental data with Eq. (\ref{model_chi}) where $\chi_\mathrm{dim}(T)$ is given by Eq. (\ref{chi_dim}) to extract the Landé factor $g$ of dimerized spins and $\chi_0$ for each field direction. The best fits are shown as solid curves in Fig. \ref{chi}(a), and the optimized parameters are $g_{ab} = 2.177(2), \chi_0 = 0$ cm$^3$mol$^{-1}_\mathrm{Cu^{2+}}$ and $g_{c^*} = 2.309(2), \chi_0 = 1.78(2)\times10^{-4}$ cm$^3$mol$^{-1}_\mathrm{Cu^{2+}}$ for $\boldsymbol{H}_{ab}$ and $\boldsymbol{H}_{c^*}$, respectively.

To reaffirm the optimized parameters, we further perform QMC simulations to calculate the reduced magnetization, $M^*(h)$, with the same set of coupling parameters for $0 \le h \le 7\times10^{-2}$, where $h = g\mu_\mathrm{B}H/J_\mathrm{max}$ and $T = 120$ K using the \textsc{directed-loop} \cite{Syljuasen2002, Syljuasen2003} algorithm. The calculation conditions are the same as for $\chi^*(t)$. $M_\mathrm{dim}(H)$ is related to $M^*(h)$ by
\begin{align*}
   M_\mathrm{dim}(H) = {}& g\mu_\mathrm{B}M^*\left(\frac{g\mu_\mathrm{B}H}{J_\mathrm{max}}\right).
\end{align*}
The simulated $M_\mathrm{dim}(H)$ is calculated with the optimized Landé factor $g$ and then compared with the background-corrected experimental data. As seen in Figs. \ref{chi}(b) and \ref{chi}(c), a reasonable agreement is achieved between the data and model calculations confirming the validity of the Landé factors and the exchange couplings.

\subsection{Triplet excitations}

\begin{figure*}
	\includegraphics[width=\textwidth]{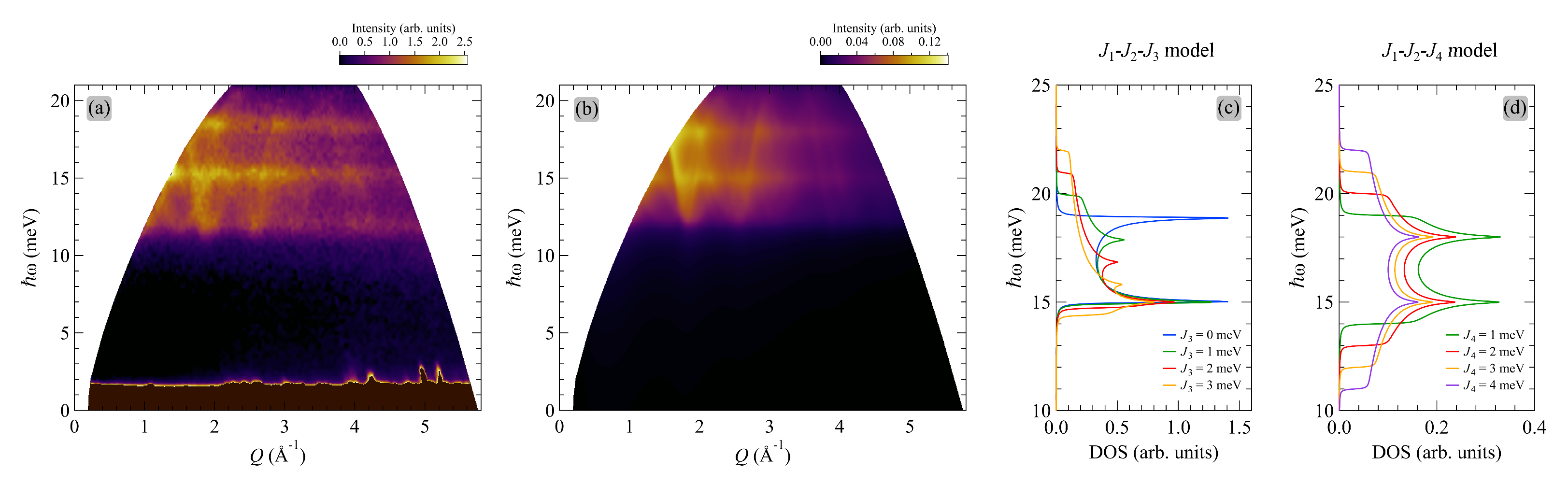}
	\caption{(a) The observed powder-averaged magnetic intensity map at $T = 4$ K (b) The calculated powder-averaged magnetic intensity map using the $J_1$-$J_2$-$J_4$ model with $J_1 = 16.5$ meV, $J_2 = -6$ meV, and $J_4 = 3$ meV. The density of states (DOS) of (c) the $J_1$-$J_2$-$J_3$ model and of (d) the $J_1$-$J_2$-$J_4$ model.}
	\label{amr}
\end{figure*}

\begin{figure}
	\includegraphics[width = 0.75\columnwidth]{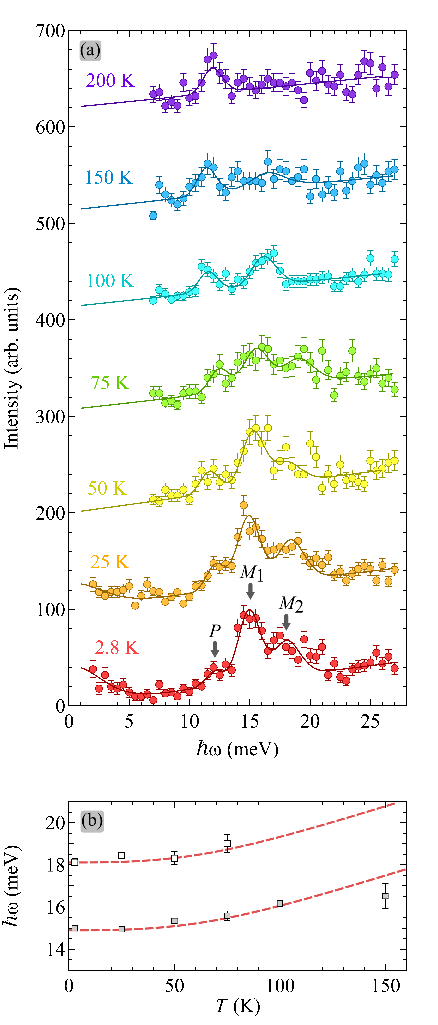}
	\caption{(a) $E$-scan data with $Q = 1.97\;\mathrm{\AA}^{-1}$ at 2.8 K. Solid lines are best fits to the model described in text. (b) Temperature dependence of $\hbar\omega$ for $M_1$ (solid squares) and $M_2$ (open squares) peaks. Dashed lines are guides to the eyes. Error bars represent one standard deviation.}
	\label{bt7}
\end{figure}

Data reduction and powder averaging of the TOF INS data collected from the AMATERAS spectrometer were performed using the \textsc{utsusemi} software package \cite{Inamura2013}. The obtained intensity maps at $T = 4$ and 200~K are displayed in Figs. S.1(a) and S.1(b), respectively, in the Supplemental Material \cite{SuppMat}. Also included in Fig. S.1(a) in the Supplemental Material \cite{SuppMat} is the excitation energy (circle symbols) extracted from the $E$-scan data taken from the BT-7 spectrometer at $T = 2.8$~K \cite{SuppMat}. In Fig. S.1(b) in the Supplemental Material \cite{SuppMat}, at $T = 200$~K where the magnetic correlations are expected to subside (see also Fig. \ref{bt7}(a)), there remain excitations whose intensity noticeably increases with the magnitude of the wave vector $Q$ and temperature. We, therefore, attribute these excitations to phonons. To obtain the magnetic scattering intensity map, $I_\mathrm{mag}(Q,\omega)$, as shown in Fig. \ref{amr}(a), we subtracted the data at the two temperatures after rescaling the data at 200 K by the temperature-dependent factor $1 + n(T) = (1 - \exp(-E/k_\mathrm{B}T))^{-1}$, where $n(T)$ is the Bose factor. It should be noted that the scattering intensity seen at $\hbar\omega = 12$ meV and $Q > 3$ \AA$^{-1}$ is most likely due to residual phonon intensity. To acquire the key characteristics of the magnetic scattering, $I_\mathrm{mag}(Q,\omega)$ is then integrated over the wave vector $1\;\mathrm{\AA^{-1}} \le Q \le 3\;\mathrm{\AA^{-1}}$. The integrated intensity shown in Fig. S.2(b) in the Supplemental Material \cite{SuppMat} features two peaks at $\hbar\omega = 15$ and 18 meV and covers an extensive range from the band bottom at $\sim$~11 meV to the band top at $\sim$~21 meV amounting to a sizable bandwidth of $\sim$~10 meV \cite{SuppMat}. No magnetic excitation is detected at $\hbar\omega \lesssim 11$~meV or $\hbar\omega \gtrsim 21$~meV.

To explore the temperature evolution of the excitation spectrum, we next examine the TA $E$-scan data collected at base and elevated temperatures from the BT-7 spectrometer. Figure \ref{bt7}(a) shows the data measured at $2.8\;\mathrm{K} \le T \le 200\;\mathrm{K}$ with a fixed wave vector $Q = 1.96\;\mathrm{\AA}^{-1}$. At $T = 2.8$~K, the data exhibit three excitation peaks at 12, 15, and 18 meV, denoted by $P$, $M_1$, and $M_2$, respectively. At elevated temperatures, the $P$ peak becomes stronger whereas the $M_1$ and $M_2$ peaks weaken and eventually disappear at 200 K. Therefore, the $M_1$ and $M_2$ peaks are consistent with the two peaks observed in the integrated magnetic intensity shown in Fig. S.2(b) in the Supplemental Material \cite{SuppMat}. For the $P$ peak, however, it is essential to note that in the TOF data (see Figs. S.1(a) and S.1(b) in the Supplemental Material \cite{SuppMat}) there exist a phonon band as well as magnetic scattering at $\hbar\omega = 12$~meV. Therefore, the $P$ peak, especially at $T = 2.8$~K, results from the superposition of these two sources. As the temperature is elevated, the magnetic contribution diminishes leaving only the enhanced phonon contribution. We fit the spectra to a Lorentzian function combined with a linear background to acquire the excitation energy, $\hbar\omega$, for each $T$ and $Q$. The width of the $M_1$ and $M_2$ peaks are assumed identical. Fitting the data (i) at $T = 2.8$~K and with $1.12\;\mathrm{\AA}^{-1} \le Q \le 3.09\;\mathrm{\AA}^{-1}$ and (ii) at $2.8\;\mathrm{K} \le T \le 200\;\mathrm{K}$ with $Q = 1.96\;\mathrm{\AA}^{-1}$ yields $\hbar\omega$ plotted as circle symbols in Fig. S.1(a) in the Supplemental Material \cite{SuppMat} and a $T$-dependent $\hbar\omega$ shown in Fig. \ref{bt7}(b). In alignment with $\chi(T)$ data, the monotonic increase in $\hbar\omega(T)$ points out that the gapped excitation observed in Cu$_5$SbO$_6$ is a triplet excitation (or triplon) since, in a dimerized magnet, triplons are subjected to the hard-core constraint that prevents them from occupying the same dimer site resulting in repulsion between them \cite{Xu2007, Essler2008, Exius2010, Normand2011} as opposed to long-range ordered magnets for which the gap decreases with increasing temperature due to the reduction in ordered moment.

\begin{figure*}
	\includegraphics[width = \textwidth]{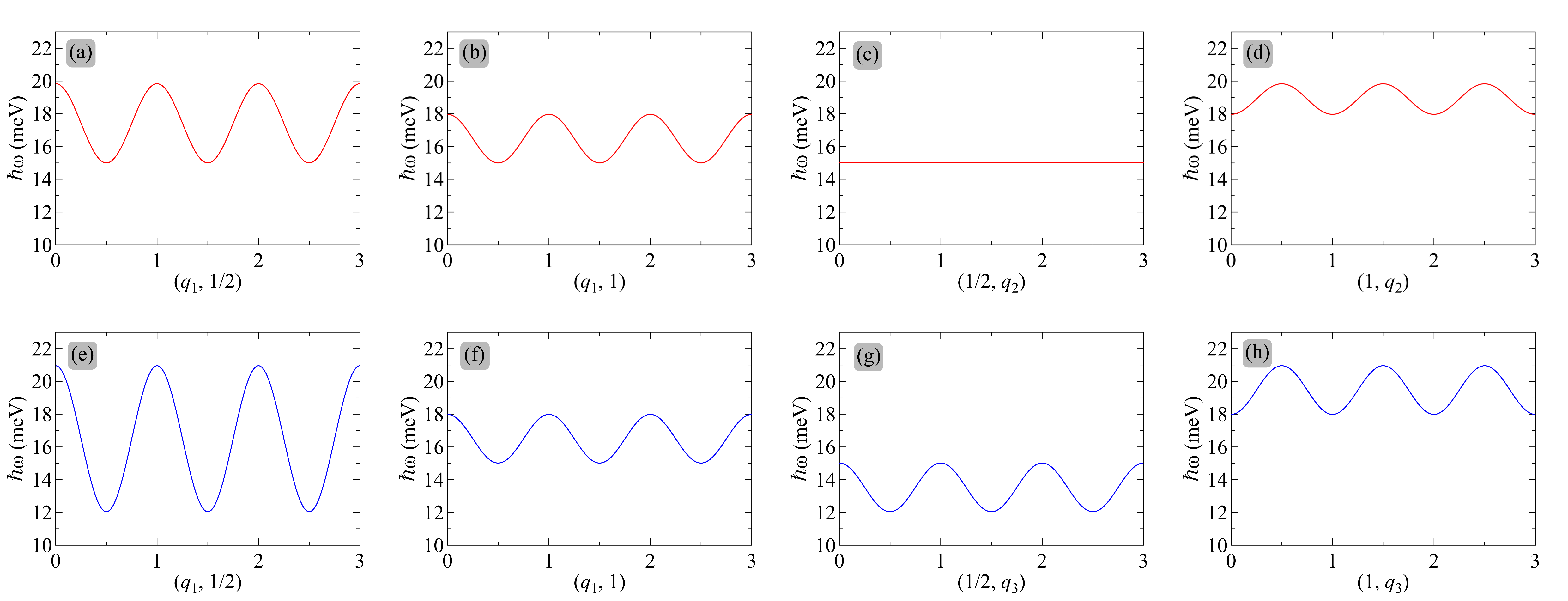}
	\caption{(a)-(d) Dispersions along $q_1$ and $q_2$ directions for the $J_1$-$J_2$-$J_3$ model calculated with $J_1 = 16.95$ meV, $J_2 = -3.9$ meV, and $J_3 = 0.9$ meV. (e)-(h) Dispersions along $q_1$ and $q_3$ directions for the $J_1$-$J_2$-$J_4$ model calculated with $J_1 = 16.5$ meV, $J_2 = -6$ meV, and $J_4 = 3$ meV.}
	\label{disp_plots}
\end{figure*}

To acquire an appropriate spin Hamiltonian for Cu$_5$SbO$_6$, we employ the first-order dimer expansion to derive the energy dispersion, $\Delta E_{\boldsymbol{Q}}$, of triplet excitations for both $J_1$-$J_2$-$J_3$ and $J_1$-$J_2$-$J_4$ models (a detailed derivation can be found in the Appendix). The expressions for the dispersion of both models are given in Eqs. (\ref{disp_1}) and (\ref{disp_2}). In these expressions, the wave vector $\boldsymbol{Q}$ is written in terms of the basis vectors $\boldsymbol{b}_1$, $\boldsymbol{b}_2$, and $\boldsymbol{b}_3$ defined in Eqs. (\ref{rec_dim_latt_1})-(\ref{rec_dim_latt_3}). Here, $q_i$ represents the component of each basis vector. Let us begin with the $J_1$-$J_2$-$J_3$ model with the dispersion given in Eq. (\ref{disp_1}). For an isolated FM-AFM chain, the dispersion along the $q_1$ direction oscillates around the energy set by the intradimer AFM $J_1$ coupling with the amplitude determined by the magnitude of the interdimer FM $J_2$ coupling. The incorporation of the interchain AFM $J_3$ coupling predominantly only results in a modulation of the top of the dispersion along the $q_2$ direction (see Figs. \ref{disp_plots}(a)-\ref{disp_plots}(d)). To see how the dispersion produces the $M_1$ and $M_2$ peaks in the magnetic scattering, we consider the density of states (DOS), $D(\omega)$, as defined by
\begin{align*}
	D(\omega) \propto \int d\boldsymbol{Q}\;\delta(\hbar\omega - \Delta E_{\boldsymbol{Q}}),
\end{align*}
where the integration is over a unit cell of the reciprocal lattice formed by dimers. Approximating the delta function by a Lorentzian function, $\delta(x) = \epsilon/\pi(x^2 + \epsilon^2)$, where $\epsilon$ is a small positive number, we finally obtain DOS for the $J_1$-$J_2$-$J_3$ model as shown in Fig. \ref{amr}(c). The result is presented for four different values of the interchain coupling $J_3$ with $J_1 = 16.95$ meV, $J_2 = -3.9$ meV. When $J_3 = 0$, two peaks appear at the top and bottom of the DOS. As $J_3$ increases, the peak at the top shifts toward lower energy and the tail develops toward higher energy. Although $J_3 \approx 1$ meV correctly reproduces peaks in DOS at 15 and 18 meV, this model does not provide a tail below 15 meV regardless of the value of $J_3$, which is in contradiction to the observed spectrum. Therefore, the $J_1$-$J_2$-$J_3$ model is not viable in explaining the data.

We next turn to the $J_1$-$J_2$-$J_4$ model. The introduction the interlayer AFM $J_4$ coupling to the system of isolated FM-AFM chains delivers a slightly distinct effect on the dispersion. Unlike with $J_3$, it is found that modulation along the $q_3$ direction appears at both the top and the bottom of the dispersion (see Figs. \ref{disp_plots}(e)-\ref{disp_plots}(h)). To obtain DOS for this model, we first consider the dispersion in Eq. (\ref{disp_2}) from which the energy of $M_1$ and $M_2$ peaks, denoted by $E_1$ and $E_2$, respectively, are given by
\begin{align*}
	E_1 = {}& J_1 - \frac{|J_2|}{2} + \frac{|J_4|}{2}, \\
	E_2 = {}& J_1 + \frac{|J_2|}{2} - \frac{|J_4|}{2},
\end{align*}
which give $2J_1 = E_1 + E_2$ and $|J_2| = |J_4| + (E_2 - E_1)$. Using $E_1 = 15.0(1)$ meV and $E_2 = 18.0(2)$ meV, we can identify $J_1 = 16.5(1)$ meV and $|J_2| = |J_4| + 3.0$ meV. Figure \ref{amr}(d) shows DOS of the $J_1$-$J_2$-$J_4$ model calculated with four different values of $|J_4|$. With these coupling parameters the DOS correctly reproduces peaks at 15 and 18 meV. Moreover, in contrast to the previous model, as $J_4$ increases, tails in DOS develop at both higher and lower energy, which resembles the pattern observed in the INS spectrum. Based on the INS data, one can obtain an estimate of $|J_4| \approx 3$ meV, and, thus, $|J_2| \approx 6$ meV. It is essential to note that in the presence of stacking faults, although the magnitude of $J_4$ might be unchanged as the path responsible for the coupling still involves the same ions and geometry, the dimer connectivity is different and the $J_1$-$J_2$-$J_4$ model might be inapplicable.

It is natural to extend the model to include the $J_3$ coupling in the $J_1$-$J_2$-$J_4$ model. To this end, let us examine the effect of $J_3$ on the current model by calculating the DOS of the extended model with different values of $J_3$ as described in the Supplemental Material \cite{SuppMat}. As can be seen in Fig. S.2(c) in the Supplemental Material \cite{SuppMat}, the presence of the $J_3$ coupling results in a suppression of the 18 meV peak and a further elongation of the tail above 21~meV. These effects become more enhanced as $J_3$ grows. Comparison with the integrated magnetic intensity in Fig. S.2(b) in the Supplemental Material \cite{SuppMat} suggests an upper bound for $J_3$ of merely $\sim0.3$~meV, which can safely be neglected when taking the uncertainty of the coupling parameters into account. Finally, to verify the $J_1$-$J_2$-$J_4$ model, a powder-averaged INS spectrum was simulated using the proposed set of coupling parameters as outlined in the Supplemental Material \cite{SuppMat}. The simulated result shown in Fig. \ref{amr}(b) compares very well with the observed spectrum justifying the $J_1$-$J_2$-$J_4$ model as adequate and appropriate for Cu$_5$SbO$_6$.

Future INS studies on high-quality single-crystal samples may shed more light on some exotic aspects of the triplon excitations and allow the spin-spin correlations to be mapped out more precisely. The latter, as discussed in Refs. \cite{Verstraete2003, Jin2004, Brukner2006}, play a crucial role in determining the degree of hidden quantum entanglement in a many-body system such as a spin chain. It has been argued in these studies that the spatial extent of quantum entanglement is directly related to the spin-spin correlations that can be measured experimentally by means of the neutron scattering technique. In this arena, as pointed out in Ref. \cite{Stone2007}, an alternating FM-AFM Heisenberg spin chain may triumph over its alternating AFM-AFM counterpart owing to the nonzero spin-spin correlations on both FM and AFM bonds. Given this, Cu$_5$SbO$_6$ can serve as a model system for exploring the spatial evolution of quantum entanglement manifesting in a many-body system.

\section{Conclusion}

We employed x-ray spectroscopy and electron microscopy to refine the crystal structure and the stacking-fault model of Cu$_5$SbO$_6$. Based on the crystal structure, Cu$_5$SbO$_6$ can accommodate alternating FM-AFM dimerized spin chains similar to the related compounds with the FM bond forming between Cu$^{2+}$ ions within the Cu$_2$O$_6$ double plaquettes and the AFM bond, hence, the dimers, forming across the hexagon in the honeycomb planes. However, in contrast to the related compounds, the interchain couplings that arise between the honeycomb planes can be significant for Cu$_5$SbO$_6$ owing to the different geometry of the superexchange paths that enhance orbital overlap. Stacking faults present in powder samples are a result of two inequivalent in-plane translations, i.e., (1/3, 0, 0) and (2/3, 0, 0), that interchangeably shift the honeycomb planes along the $a$ direction with a frequency depending on the synthesis temperature.

Based on a first-order dimer expansion approximation, we showed that, unlike the related compounds, the $J_1$-$J_2$-$J_3$ model cannot be a complete description for the magnetic properties of Cu$_5$SbO$_6$. With a specific set of coupling parameters, the density of states of such a model, whereas containing peaks at 15 and 18 meV, does not exhibit a tail below 15 meV in the INS spectrum. On the other hand, the $J_1$-$J_2$-$J_4$ model is capable of reproducing the entire spectrum and, therefore, proves to be a more appropriate starting point for the spin Hamiltonian for Cu$_5$SbO$_6$. The coupling parameters of the $J_1$-$J_2$-$J_4$ model estimated from the powder-averaged INS data are $J_1 = 16.5(1)$ meV, $|J_2| \approx 6\;\mathrm{meV}$, and $|J_4| \approx 3\;\mathrm{meV}$. With the obtained coupling parameters, we further tested the model by simulating a temperature-dependent susceptibility and field-dependent magnetization. Fitting the simulated susceptibility to the experimental data returns sensible Landé factors, i.e., $g_{ab} = 2.177(2)$ and $g_{c^*} = 2.309(2)$, for Cu$^{2+}$ moments in plaquette environments. Moreover, the simulated magnetization curves calculated with the optimized Landé factors compare reasonably well with the experimental data. Considering these results, Cu$_5$SbO$_6$ may be included in a rare collection of compounds that host interacting FM-AFM quantum spin chains, which can serve as a playground for exploring emergent quantum collective phenomena.

\begin{acknowledgments}

This research has received funding support from Mahidol University (Strategic Research Fund) fiscal year 2024. C.P. was supported by the NSRF via the Program Management Unit for Human Resources \& Institutional Development, Research and Innovation (PMU-B) (Grant No. B13F680079). C.P. is grateful for the support by the DPST scholarship from the Institute for the Promotion of Teaching Science and Technology. C.P. and K. Matan thank Dr. Khetpakorn Chakarawet from the Department of Chemistry, Mahidol University, for his assistance in single-crystal XRD measurements. C.P. and K. Matan thank Masaki Ageishi for his assistance in TEM experiments at Tohoku University, Japan. W.-T.C. acknowledges the support of the National Science and Technology Council in Taiwan (Grants No. 113-2124-M-002-006 and No. 114-2124-M-002-010) and the Featured Research Center Program within the framework of the Higher Education Sprout Project by the Ministry of Education in Taiwan (Grant No. 113L9008). K. Morita was supported by JSPS KAKENHI (Grant No. JP25K07230). The experiment on AMATERAS was performed with the approval of J-PARC (Proposal No. 2019A0194 (R)). We are grateful for the support from the National Institute of Standards and Technology, U.S. Department of Commerce, for providing neutron research facilities used in this work. The identification of any commercial product or trade name does not imply endorsement or recommendation by the National Institute of Standards and Technology. Synchrotron Light Research Institute (Public Organization), Thailand, is acknowledged for the provision of XAS beamtime.

\end{acknowledgments}

\appendix*

\section{Dimer expansion for triplet excitations}

\begin{figure}
	\includegraphics[width = 0.8\columnwidth]{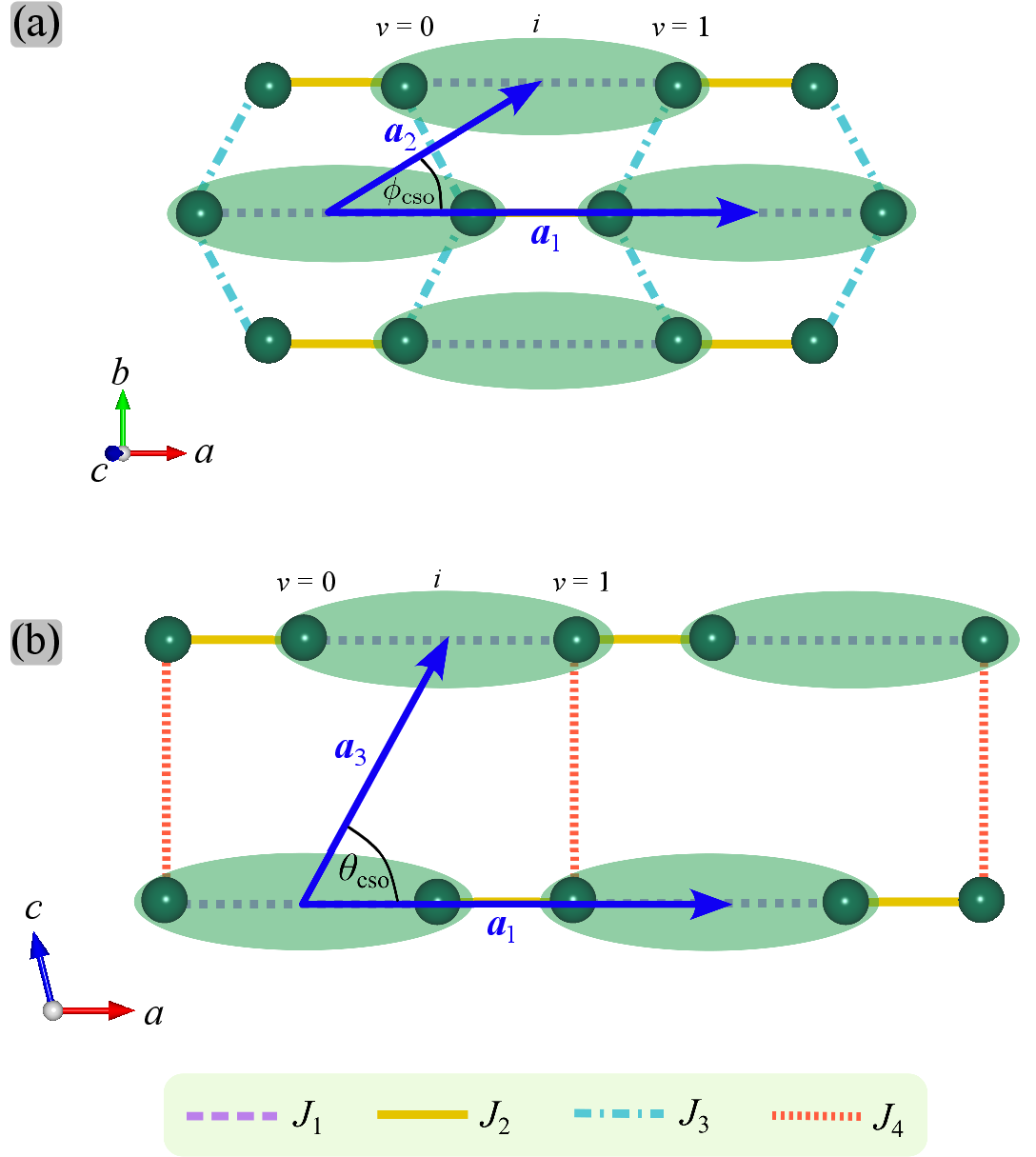}
	\caption{The lattice formed by dimers depicted as green ovals for the $J_1$-$J_2$-$J_3$ and $J_1$-$J_2$-$J_4$ models are shown in panels (a) and (b), respectively. $i$ is the dimer identifier in the lattice, and $\nu = 0, 1$ is the spin identifier in the dimer. $\boldsymbol{a}_1$, $\boldsymbol{a}_2$, and $\boldsymbol{a}_3$ are the primitive translation vectors.}
	\label{dimer_lattice}
\end{figure}

In this Appendix, the dispersion relation as well as the dynamical structure factor for the $J_1$-$J_2$-$J_3$ and $J_1$-$J_2$-$J_4$ models are derived. Figure \ref{dimer_lattice} shows the lattices formed by dimers for the two models where the primitive translation vectors $\boldsymbol{a}_1$, $\boldsymbol{a}_2$, $\boldsymbol{a}_3$ are expressed as
\begin{align}
	\boldsymbol{a}_1 = {}& l_1\boldsymbol{e}_x, \\
	\boldsymbol{a}_2 = {}& l_2(\cos\phi_\mathrm{cso}\boldsymbol{e}_x + \sin\phi_\mathrm{cso}\boldsymbol{e}_y), \\
	\boldsymbol{a}_3 = {}& l_3(\cos\theta_\mathrm{cso}\boldsymbol{e}_x + \sin\theta_\mathrm{cso}\boldsymbol{e}_z).
	\label{dim_latt}
\end{align}
$\boldsymbol{a}_1$ and $\boldsymbol{a}_2$ form the dimer lattice of the $J_1$-$J_2$-$J_3$ model whereas $\boldsymbol{a}_1$ and $\boldsymbol{a}_3$ form that of the $J_1$-$J_2$-$J_4$ model. The primitive translation vectors of the reciprocal dimer lattice $\boldsymbol{b}_1$, $\boldsymbol{b}_2$, $\boldsymbol{b}_3$ are
\begin{align}
	\boldsymbol{b}_1 = {}& \frac{2\pi}{l_1}(\boldsymbol{e}_x - \cot\phi_\mathrm{cso}\boldsymbol{e}_y - \cot\theta_\mathrm{cso}\boldsymbol{e}_z),
	\label{rec_dim_latt_1} \\
	\boldsymbol{b}_2 = {}& \frac{2\pi}{l_2}\csc\phi_\mathrm{cso}\boldsymbol{e}_y,
	\label{rec_dim_latt_2} \\
	\boldsymbol{b}_3 = {}& \frac{2\pi}{l_3}\csc\theta_\mathrm{cso}\boldsymbol{e}_z.
	\label{rec_dim_latt_3}
\end{align}
$\boldsymbol{b}_1$ and $\boldsymbol{b}_2$ form the reciprocal dimer lattice of the $J_1$-$J_2$-$J_3$ model whereas $\boldsymbol{b}_1$ and $\boldsymbol{b}_3$ form that of the $J_1$-$J_2$-$J_4$ model. Note that for Cu$_5$SbO$_6$, $l_1 = 9.00$~\AA, $l_2 = 5.30$~\AA, $l_3 = 6.55$~\AA, $\theta_\mathrm{cso} = 62^\circ$, and $\phi_\mathrm{cso} = 30^\circ$.

\subsection{Energy dispersion}

In our Hamiltonian $H = H_0 + H'$ for the two models, the common unperturbed term $H_0$ is defined by
\begin{align*}
	H_0 = J_1\sum_{i}\boldsymbol{S}_{i,0}\cdot\boldsymbol{S}_{i,1},
\end{align*}
where $i$ is an identifier given to a dimer and $\boldsymbol{S}_{i,\nu}$ is the $\nu$th spin operator in the $i$th dimer (see Fig. \ref{dimer_lattice}). The perturbation term $H'$ for each model is given by
\begin{align*}
	H_{123}' = {}& J_2\sum_{i}\boldsymbol{S}_{i,1}\cdot\boldsymbol{S}_{i + \boldsymbol{a}_1,0} \\
	   + {}& J_3\sum_{i}\boldsymbol{S}_{i,1}\cdot(\boldsymbol{S}_{i + \boldsymbol{a}_2,0} + \boldsymbol{S}_{i + \boldsymbol{a}_1 - \boldsymbol{a}_2,0}), \\
	   = {} & \sum_i\sum_{\rho = \boldsymbol{a}_1,\boldsymbol{a}_2, \boldsymbol{a}_1 - \boldsymbol{a}_2}J_{\rho}\boldsymbol{S}_{i,1}\cdot\boldsymbol{S}_{i + \rho,0} \\
	H_{124}' = {}& J_2\sum_{i}\boldsymbol{S}_{i,1}\cdot\boldsymbol{S}_{i + \boldsymbol{a}_1,0} \\
	   + {}& J_4\sum_{i}\boldsymbol{S}_{i,1}\cdot\boldsymbol{S}_{i + \boldsymbol{a}_1 - \boldsymbol{a}_3,0}, \\
	   = {} & \sum_i\sum_{\rho = \boldsymbol{a}_1, \boldsymbol{a}_1 - \boldsymbol{a}_3}J_{\rho}\boldsymbol{S}_{i,1}\cdot\boldsymbol{S}_{i + \rho,0}.
\end{align*}

The wave function of the singlet sea $|s\rangle$ is expressed as
\begin{align*}
	|s\rangle = \prod_{n} |s\rangle_n,\;\mathrm{where}\;|s\rangle_n \equiv \frac{1}{\sqrt{2}}(|\uparrow\downarrow\rangle_n - |\downarrow\uparrow\rangle_n).
\end{align*}
Here, $|\uparrow\downarrow\rangle_n$, $|\downarrow\uparrow\rangle_n$, ..., represent the state of the $n$th dimer in the Ising basis. We also define the single-triplet-dimer state $|t^+_l\rangle$ (STD), in which only the $l$th dimer in the singlet sea is in $|t^+\rangle_l \equiv |\uparrow\uparrow\rangle_l$, as
\begin{align*}
	|t^+_l\rangle = |t^+\rangle_l\prod_{n(\neq l)}|s\rangle_n.
\end{align*}
Multiplying $H_0$ by $|t^+_l\rangle$, we immediately get
\begin{align*}
	H_0|t^+_l\rangle = J_1\sum_{i}(\boldsymbol{S}_{i,0}\cdot\boldsymbol{S}_{i,1})|t^+_l\rangle = J_1\left(-\frac{3}{4}N + 1\right)|t^+_l\rangle.
\end{align*}
Next, we multiply $H'$ by $|t^+_l\rangle$. Notice that each exchange operator in the perturbation term involves two dimers. In order to evaluate $H'|t^+_l\rangle$, we explicitly calculate the following three patterns of terms for the two dimers. The first one is $\boldsymbol{S}_{l,1}\cdot\boldsymbol{S}_{l+\rho,0}|t^+\rangle_l|s\rangle_{l+\rho}$, which can be evaluated as
\begin{widetext}
\begin{align*}
	\boldsymbol{S}_{l,1}\cdot\boldsymbol{S}_{l+\rho,0}|t^+\rangle_l|s\rangle_{l+\rho} = {}& \left[S^z_{l,1}S^z_{l+\rho,0} + \frac{1}{2}\left(\underbrace{S^+_{l,1}}_{\to 0}S^-_{l+\rho,0} + S^-_{l,1}S^+_{l+\rho,0}\right)\right]|t^+\rangle_l|s\rangle_{l+\rho} \\
	= {}& \frac{1}{4}|t^+\rangle_l|t^0\rangle_{l+\rho} - \frac{1}{2}\frac{1}{\sqrt{2}}\left(|t^0\rangle_l + |s\rangle_l\right)\frac{1}{\sqrt{2}}|t^+\rangle_{l+\rho} \\
	= {}& -\frac{1}{4}|s\rangle_l|t^+\rangle_{l+\rho} + \frac{1}{4}|t^+\rangle_l|t^0\rangle_{l+\rho} - \frac{1}{4}|t^0\rangle_l|t^+\rangle_{l+\rho},
\end{align*}
\end{widetext}
which gives
\begin{align*}
	\boldsymbol{S}_{l,1}\cdot\boldsymbol{S}_{l+\rho,0}|t^+_l\rangle = -\frac{1}{4}|t^+_{l+\rho}\rangle + \textnormal{(non-STD states)}.
\end{align*}
The second one is $\boldsymbol{S}_{l-\rho,1}\cdot\boldsymbol{S}_{l,0}|s\rangle_{l-\rho}|t^+\rangle_{l}$, which can be evaluated as
\begin{align*}
	\boldsymbol{S}_{l-\rho,1}\cdot\boldsymbol{S}_{l,0}|s\rangle_{l-\rho}|t^+\rangle_{l} = &-\frac{1}{4}|t^+\rangle_{l-\rho}|s\rangle_{l} - \frac{1}{4}|t^0\rangle_{l-\rho}|t^+\rangle_l \\
	& + \frac{1}{4}|t^+\rangle_{l-\rho}|t^0\rangle_l,
\end{align*}
which gives
\begin{align*}
	\boldsymbol{S}_{l-\rho,1}\cdot\boldsymbol{S}_{l,0}|t^+_l\rangle = -\frac{1}{4}|t^+_{l-\rho}\rangle + \textnormal{(non-STD states)}.
\end{align*}
The final one is $\boldsymbol{S}_{i,1}\cdot\boldsymbol{S}_{i+\rho,0}|s\rangle_{i}|s\rangle_{i+\rho}$ with $i \ne l$ and $i + \rho \ne l$, which can be evaluated as
\begin{align*}
	\boldsymbol{S}_{i,1}\cdot\boldsymbol{S}_{i+\rho,0}|s\rangle_{i}|s\rangle_{i+\rho} = &-\frac{1}{4}|t^0\rangle_i|t^0\rangle_{i+\rho} + \frac{1}{4}|t^+\rangle_i|t^-\rangle_{i+\rho} \\
	& - \frac{1}{4}|t^-\rangle_i|t^+\rangle_{i+\rho},
\end{align*}
which gives
\begin{align*}
	\boldsymbol{S}_{i,1}\cdot\boldsymbol{S}_{i+\rho,0}|t^+_l\rangle = \textnormal{(non-STD states)}.
\end{align*}
Using the above results, $H'|t^+_l\rangle$ can be evaluated as follows:
\begin{align*}
	H'|t^+_l\rangle = -\frac{1}{4}\sum_{\rho}J_{\rho}(|t^+_{l+\rho}\rangle + |t^+_{l-\rho}\rangle) + \textnormal{(non-STD states)}.
\end{align*}
Let $H_\mathrm{eff}$ be the effective Hamiltonian of the STD states, in which the energy is measured from the ground-state energy,
\begin{align*}
	E_\mathrm{g} = -\frac{3}{4}J_1N + \textnormal{(second-order or higher terms)}.
\end{align*}
Then, the matrix elements of $H_\mathrm{eff}$ within the first-order perturbation for each model are found to be
\begin{align*}
\langle t^+_{l'}|H_\mathrm{eff}^{123}|t^+_l\rangle \simeq {}& \langle t^+_{l'}|H_0|t^+_l\rangle + \langle t^+_{l'}|H'_{123}|t^+_l\rangle - E_\mathrm{g}\delta_{l',l} \\
= {}& J_1\delta_{l',l} - \frac{1}{4}\sum_{\rho}J_{\rho}(\delta_{l',l+\rho} + \delta_{l',l-\rho}) \\
= {}& J_1\delta_{l',l} - \frac{J_2}{4}(\delta_{l',l+\boldsymbol{a}_1} + \delta_{l',l-\boldsymbol{a}_1}) \notag \\
 - \frac{J_3}{4}(\delta_{l',l+\boldsymbol{a}_2} + &{} \delta_{l',l-\boldsymbol{a}_2} + \delta_{l',l+\boldsymbol{a}_1 - \boldsymbol{a}_2} + \delta_{l',l-\boldsymbol{a}_1+\boldsymbol{a}_2}), \\
 \langle t^+_{l'}|H_\mathrm{eff}^{124}|t^+_l\rangle \simeq {}& \langle t^+_{l'}|H_0|t^+_l\rangle + \langle t^+_{l'}|H'_{124}|t^+_l\rangle - E_\mathrm{g}\delta_{l',l} \\
= {}& J_1\delta_{l',l} - \frac{1}{4}\sum_{\rho}J_{\rho}(\delta_{l',l+\rho} + \delta_{l',l-\rho}) \\
= {}& J_1\delta_{l',l} - \frac{J_2}{4}(\delta_{l',l+\boldsymbol{a}_1} + \delta_{l',l-\boldsymbol{a}_1}) \notag \\
 & - \frac{J_4}{4}(\delta_{l',l+\boldsymbol{a}_1 - \boldsymbol{a}_3} + \delta_{l',l-\boldsymbol{a}_1+\boldsymbol{a}_3}).
\end{align*}

To diagonalize $H_\mathrm{eff}$, we introduce a Fourier transform:
\begin{align*}
	|t^+_{\boldsymbol{Q}}\rangle = \frac{1}{\sqrt{N}}\sum_i|t^+_i\rangle e^{-i\boldsymbol{Q}\cdot\boldsymbol{r}_i},
\end{align*}
where $N$ is the total number of dimers, and $\boldsymbol{r}_i$ is the position vector of the $i$th dimer. This transformation diagonalizes the effective Hamiltonian and gives the triplet excitation spectrum $\Delta E_{\boldsymbol{Q}} \equiv \langle t^+_{\boldsymbol{Q}}|H_\mathrm{eff}|t^+_{\boldsymbol{Q}}\rangle$:
\begin{align}
	\Delta E_{\boldsymbol{Q}}^{123} = J_1 {}& - \frac{J_2}{2}\cos(\boldsymbol{Q}\cdot\boldsymbol{a}_1)\notag \\
	{}& - J_3\cos\left(\frac{\boldsymbol{Q}\cdot\boldsymbol{a}_1}{2}\right)\cos\left[\boldsymbol{Q}\cdot\left(\boldsymbol{a}_2 - \frac{1}{2}\boldsymbol{a}_1\right)\right].
	\label{disp_01}
\end{align}
Similarly, for the $J_1$-$J_2$-$J_4$ model,
\begin{align}
	\Delta E_{\boldsymbol{Q}}^{124} = J_1 - \frac{J_2}{2}\cos(\boldsymbol{Q}\cdot\boldsymbol{a}_1) - \frac{J_4}{2}\cos[\boldsymbol{Q}\cdot(\boldsymbol{a}_3 - \boldsymbol{a}_1)].
	\label{disp_02}
\end{align}
Defining $\boldsymbol{Q} = q_1\boldsymbol{b}_1 + q_2\boldsymbol{b}_2 + q_3\boldsymbol{b}_3$, we then have 
\begin{align}
	\boldsymbol{Q}\cdot\boldsymbol{a}_j = \sum_i q_i(\boldsymbol{b}_i\cdot\boldsymbol{a}_j) = \sum_i q_i(2\pi\delta_{ij}) = 2\pi q_j.
\end{align}
Equations (\ref{disp_01}) and (\ref{disp_02}) become
\begin{align}
	\Delta E^\mathrm{123}_{\boldsymbol{Q}} = {}& J_1 - \frac{J_2}{2}\cos(2\pi q_1) - J_3\cos(\pi q_1)\cos[\pi(2q_2 - q_1)],
	\label{disp_1} \\
	\Delta E^\mathrm{124}_{\boldsymbol{Q}} = {}& J_1 - \frac{J_2}{2}\cos(2\pi q_1) - \frac{J_4}{2}\cos[2\pi(q_3 - q_1)].
	\label{disp_2}
\end{align}

\subsection{Dynamical structure factor}

We denote the ground state by $|\Psi_\mathrm{g}\rangle$ and its energy eigenvalue by $E_\mathrm{g}$. We also denote the basis of the subspace of wave vectors $\boldsymbol{Q}$ by $\{|\lambda_{\boldsymbol{Q}}\rangle\}$. Let these be eigenstates of $H$ with energy eigenvalues $E_{\lambda\boldsymbol{Q}}$. Then, the dynamical structure factor $\mathcal{S}^{+-}(\boldsymbol{Q},\omega)$ at $T = 0$ can be written as
\begin{align*}
	\mathcal{S}^{+-}(\boldsymbol{Q},\omega) = \sum_{\lambda_{\boldsymbol{Q}}}|\langle\lambda_{\boldsymbol{Q}}|S^+_{\boldsymbol{Q}}|\Psi_\mathrm{g}\rangle|^2\delta(\omega - E_{\lambda_{\boldsymbol{Q}}} + E_\mathrm{g}).
\end{align*}
The spin operator $S^+_{\boldsymbol{Q}}$ on the right-hand side is defined by
\begin{align*}
	S^+_{\boldsymbol{Q}} = \frac{1}{\sqrt{2N}}\sum_{i,\nu}e^{i\boldsymbol{Q}\cdot\left(\boldsymbol{r}_i+\boldsymbol{d}_{\nu}\right)}S^+_{i,\nu},
\end{align*}
where $2N$ represents the total number of spins, $\boldsymbol{r}_i$ represents the position of the $i$-th dimer, and $\boldsymbol{r}_i+\boldsymbol{d}_{\nu}$ represents the position of the spin $\boldsymbol{S}_{i,\nu}$.

Let the ground state $|\Psi_\mathrm{g}\rangle = |s\rangle$. Then, we need to evaluate
\begin{align}
	S^+_{\boldsymbol{Q}}|s\rangle = \frac{1}{\sqrt{2N}}\sum_{i,\nu}e^{i\boldsymbol{Q}\cdot\left(\boldsymbol{r}_i+\boldsymbol{d}_{\nu}\right)}S^+_{i,\nu}|s\rangle.
	\label{spin_op}
\end{align}
Noting that
\begin{align*}
	S^+_{i,\nu}|s\rangle = {}& S^+_{i,\nu}|s\rangle_i\prod_{j(\ne i)}|s\rangle_j \\
	= {}& (-1)^{\nu+1}|t^+\rangle_i\prod_{j(\ne i)}|s\rangle_j \\
	= {}& (-1)^{\nu+1}|t^+_i\rangle,
\end{align*}
we find that Eq.(\ref{spin_op}) can be written as
\begin{align}
	S^+_{\boldsymbol{Q}}|s\rangle = {}& \frac{1}{\sqrt{2N}}\sum_{i,\nu}e^{i\boldsymbol{Q}\cdot\left(\boldsymbol{r}_i+\boldsymbol{d}_{\nu}\right)}(-1)^{\nu+1}|t^+_i\rangle \notag \\
	= {}& \frac{1}{\sqrt{2}}\sum_{\nu}e^{i\boldsymbol{Q}\cdot\boldsymbol{d}_{\nu}}(-1)^{\nu+1}|t^+_{\boldsymbol{Q}}\rangle,
	\label{spin_op_2}
\end{align}
where
\begin{align*}
	|t^+_{\boldsymbol{Q}}\rangle = \frac{1}{\sqrt{N}}\sum_{i}|t^+_i\rangle e^{i\boldsymbol{Q}\cdot\boldsymbol{r}_i}.
\end{align*}
From Eq. (\ref{spin_op_2}), we can see that matrix elements appear only when $|\lambda_{\boldsymbol{Q}}\rangle = |t^+_{\boldsymbol{Q}}\rangle$. Therefore, $\mathcal{S}^{+-}(\boldsymbol{Q},\omega)$ can be written as
\begin{align*}
	\mathcal{S}^{+-}(\boldsymbol{Q},\omega) = |\langle t^+_{\boldsymbol{Q}}|S^+_{\boldsymbol{Q}}|s\rangle|^2\delta(\omega - \Delta E_{\boldsymbol{Q}}).
\end{align*}
The matrix element $\langle t^+_{\boldsymbol{Q}}|S^+_{\boldsymbol{Q}}|s\rangle$ is
\begin{align*}
	\langle t^+_{\boldsymbol{Q}}|S^+_{\boldsymbol{Q}}|s\rangle = \frac{1}{\sqrt{2}}\sum_{\nu}e^{i\boldsymbol{Q}\cdot\boldsymbol{d}_{\nu}}(-1)^{\nu+1}
\end{align*}
and taking the square of the absolute value, 
\begin{align*}
	|\langle t^+_{\boldsymbol{Q}}|S^+_{\boldsymbol{Q}}|s\rangle|^2 = {}& \left|\frac{1}{\sqrt{2}}\sum_{\nu}e^{i\boldsymbol{Q}\cdot\boldsymbol{d}_{\nu}}(-1)^{\nu+1}\right|^2 \notag \\
	= {}& \frac{1}{2}\left(e^{i\boldsymbol{Q}\cdot\boldsymbol{d}_{0}} - e^{i\boldsymbol{Q}\cdot\boldsymbol{d}_{1}}\right)\left(e^{-i\boldsymbol{Q}\cdot\boldsymbol{d}_{0}} - e^{-i\boldsymbol{Q}\cdot\boldsymbol{d}_{1}}\right) \notag \\
	= {}& 1 - \cos[\boldsymbol{Q}\cdot(\boldsymbol{d}_1 - \boldsymbol{d}_0)].
\end{align*}
Noting that $\boldsymbol{d}_1 - \boldsymbol{d}_0 = \frac{2}{3}\boldsymbol{a}_1$ for both models, we thus have
\begin{align*}
	\boldsymbol{Q}\cdot(\boldsymbol{d}_1 - \boldsymbol{d}_0) = \frac{4\pi}{3}q_1,
\end{align*}
and
\begin{align*}
	|\langle t^+_{\boldsymbol{Q}}|S^+_{\boldsymbol{Q}}|s \rangle|^2 = 1 - \cos\left(\frac{4\pi}{3}q_1\right).
\end{align*}
Therefore, the expression for the dynamical structure factor can be written as
\begin{align}
	\mathcal{S}^{+-}(\boldsymbol{Q},\omega) = \left[1 - \cos\left(\frac{4\pi}{3}q_1\right)\right]\delta(\omega - \Delta E_{\boldsymbol{Q}}).
\end{align}

\bibliography{cso_refs.bib}

\end{document}